\documentstyle[12pt, fleqn]{article}
\begin{document}

\title{  Relativistic Calculations for the Exclusive Photoproduction of 
$\eta$ Mesons from Complex Nuclei\thanks{
            Work supported in part by the Natural Sciences and
            Engineering Research Council of Canada} }

\author{\bf  M. Hedayati-Poor and H.S. Sherif  \\
Department of Physics, University of Alberta \\
Edmonton,  Alberta, Canada T6G 2J1}

\date{\today}

\maketitle

\begin{abstract}
A relativistic model for the quasifree photoproduction of $\eta$ meson from 
complex nuclei is developed. 
The interactions between fields are introduced through effective Lagrangians. 
Contributions from several nucleon resonances as well as nucleon Born terms 
and vector meson exchange diagrams are included. 
Nucleon and $\eta$ wave functions are solutions of Dirac and Klein-Gordon 
equations, respectively. 
Final state interactions of the outgoing particles are included via optical 
potentials. The effects of these interactions on the cross sections and 
photon asymmetries are studied and are found to be large. 
Calculations indicate that at energies near threshold the exclusive reaction 
takes place mainly through formation of the S$_{11}$( 1535 ) resonance. 
Comparisons with the non-relativistic calculations show differences 
between the two approaches both for the cross sections and photon asymmetries. 
We give some detailed predictions for the reaction observables for 
exclusive photoproduction on $^{12}$C, $^{16}$O and $^{40}$Ca.

\hspace{-.25 in}PACS number(s) 25.20.Lj, 24.10.Jv, 24.70+s, 13.60.-r, 13.60.Le

\end{abstract}

\newpage

\section{Introduction}      
\label{intro}
$\eta$ meson production reactions explore a rich domain of phenomena at 
the interface of nuclear and particle physics. 
These reactions have the potential for expanding our understanding of 
the formation of nucleon resonances and their propagation in the nuclear 
medium. 
Moreover, as has been shown recently, these reactions have allowed more 
precise determination of the mass of the $\eta$ meson and the study of its 
rare decays \cite{mass,rare}. 
The spin-isospin characteristics of the $\eta$ lead to a selectivity of 
resonance channels that can be formed through its interactions with a 
nucleon. 
At energies near the threshold of its production, there is a significant 
preference for the excitation of the S$_{11 }$(1535) resonance which is known 
to decay nearly 50\% of the time to an $\eta$ meson and a nucleon. 
By contrast, the S$_{11 }$(1650), which has identical characteristics, 
has only a decay rate below 2\%. 
The understanding of these dissimilar branching ratios is still an open 
question and may be a reflection of important and subtle differences in 
the detailed substructures of these resonances. 
The selectivity mentioned above makes the $\eta$ meson production reactions 
an important tool for studying the physics of the S$_{11 }$(1535) resonance.       

Until recently the production of $\eta$ mesons using photon beams has been 
largely confined to proton targets. 
The advent of high-duty-cycle electron accelerators such as those at the 
Jefferson Lab, 
Mainz, Bates and other laboratories, has opened up novel possibilities of 
performing production experiments on complex nuclei \cite{4pri}.

The elementary reaction $ \gamma~+~p~\rightarrow~\eta~+p$ has been the 
subject of several studies. 
Benmerrouche {\it et al.} \cite{4BEN.PRD} developed an effective
Lagrangian approach to study this reaction. 
Contributions from nucleon resonances, t-channel vector meson as well 
as the nucleon Born diagrams are included. 
Calculations were performed at the tree level and included eight free 
parameters. 
The $S_{11}(1535)$ was found to be the dominant contributor to the 
reaction at energies close to threshold. 
In addition, from comparisons with the data, the authors concluded 
that there is no clear preference for using pseudoscalar 
(~PS~) or pseudovector (~PV~) coupling forms for the interaction vertices 
of the $\eta$ meson with the nucleon and nucleon resonances. 
Bennhold and Tanabe \cite{4BEN.PLB,4BEN.NPA} developed a somewhat different 
approach to $\eta$ photoproduction on the nucleon. 
They used a coupled channel isobar model, in which the reaction is related 
to $(~\gamma ,~\pi ~)$, $(~\pi ,~\eta ~)$ and $(~\eta ,~\eta ~)$ reactions. 
They included contributions from three resonances: S$_{11 }$(1535), 
P$_{11 }$(1440) and D$_{13 }$(1520).
The data on the elementary reaction appeared to favor the PS form of the 
$\eta$NN vertex.
The authors employed the resulting elementary amplitude to study 
coherent and incoherent photoproduction of $\eta$ mesons on nuclei. 
This model for the elementary process was later extended by Tiator 
{\it et al.} \cite{4TIA.NPA} to include Born as well as $\rho$ and 
$\omega$ exchange diagrams through effective Lagrangians.  
The resulting elementary amplitudes were subsequently used to calculate the 
PWIA (Plane Wave Impulse Approximation) amplitude of the photoproduction 
reaction on light nuclei. 

The elementary amplitude mentioned above was also used by Lee {\it et al.} 
\cite{4LEE.NPA} to calculate the amplitude for quasifree photoproduction of 
$\eta$ mesons on heavier nuclei.
They performed exclusive as well as inclusive reaction calculations using 
a nonrelativistic approach. 
The initial bound nucleon is described by a harmonic oscillator wavefunction 
and final state interactions of the outgoing particles with the 
recoil nucleus are introduced through optical potentials. 
Two different optical potentials were used for the $\eta$ mesons. 
The authors found that the photon asymmetry is insensitive to the final state 
interactions of the outgoing particles as well as the size of the target 
nucleus. 

In this paper we develop a relativistic model for the (~$\gamma, ~\eta p$~) 
reaction on nuclei leading to specific final states of the residual nuclei 
( exclusive reactions ). The study of the photoproduction reactions on nuclei 
compliments the ongoing studies of the production on free nucleons. 
In addition to further investigating the reaction mechanism, 
studies on nuclei can help limit any existing ambiguities in the elementary 
process. 
Moreover since the production process is mediated by certain nucleon 
resonances and vector mesons, these studies open the possibility of 
investigating medium modifications of the properties of these hadrons in nuclei.  
As mentioned earlier, the availability of a new generation of accelerators 
capable of producing high quality beams of electrons and photons, 
has led to increased interest in eta photoproduction physics. 
Many experiments that were not possible in the past are now within reach. 

The present work is devoted to quasifree photoproduction. This type of 
reaction has the advantage of allowing measurements at small momentum 
transfer to the residual nucleus. This, in addition to enhancing the 
possibility of increased cross sections, 
compliments the information obtained from coherent and incoherent 
(~$\gamma, ~\eta $~) reactions which by necessity involve large 
momentum transfers.  

In the present study we follow a relativistic approach. 
This is motivated by the impressive successes of the relativistic mean 
field approach to nuclear dynamics. We have recently studied the 
difference between relativistic and non-relativistic treatments in 
photonuclear knockout 
\cite{HJS.NPA} and quasielastic scattering of electrons \cite{HJS.PRC}, 
and have found that there are important differences between the 
two approaches. 
In the course of the present study we shall compare our results to those of 
the work of Lee {\it et al.} \cite{4LEE.NPA} .

The main ingredients of the present treatment are as follows. 
The effective Lagrangian of Benmerrouche {\it et al.} \cite{4BEN.PRD} is used 
for the interactions between fields.  
Contributions from nucleon resonances, t-channel vector mesons as well as the 
nucleon Born diagrams are included. 
In addition to the S$_{11}$( 1535 ) resonance, we also include three 
spin-$\frac{1}{2}$ resonances ( P$_{11 }$(1440), S$_{11 }$(1650),  
P$_{11 }$(1670) ) and 
one spin-$\frac{3}{2}$ resonance ( D$_{13 }$(1520) ).
The nuclear wavefunctions are solutions of the Dirac equation with strong 
scalar and vector potentials in the spirit of the relativistic mean field 
theory of Walecka \cite{4W,4SW}. 
Calculations are carried out in the PWIA limit and also in the Distorted Wave 
Approximation (DWA) in which the final state interactions are taken into account. 
Bennhold and Wright have developed a similar model for the photoproduction 
of kaons on nuclei \cite{BL.PLB}.
A short note on the present work was reported earlier \cite{CEBAF}. 

In the following section we derive the amplitude of the A(~$\gamma $, 
$\eta$p~)A$-$1 reaction and calculate the relevant observables. 
The results of our calculations are presented in section III and our conclusions 
are given in section IV. 

\section{Formalism}
           
The model for the (~$\gamma $, $\eta$p~) reaction on nuclei has the incident 
photon interacting with a bound proton through some process denoted by the 
blob in Fig.1.
A proton and an $\eta$ meson are produced. 
The diagrams contributing to the blob of Fig.1, at the tree level, are shown 
in Fig.2. 
The Born diagrams (~s- and u-channels of the nucleon propagator~) are shown in 
Fig.(2a) and Fig.(2b),
the t-channel vector meson diagram is shown in Fig.(2c) and Figs.(2d) and (2e) 
are the nucleon resonance poles.
 
The starting point in the present approach is a relativistic interaction 
Lagrangian for a system of photons, 
nucleons and mesons from which one obtains the transition amplitude for the 
A(~$\gamma $, $\eta$p~)A$-$1 reaction. 
The amplitude then is used to calculate the observables for the reaction.

\subsection{The Interaction Lagrangian}
In the photoproduction of $\eta$ mesons from complex nuclei, the reaction is taking 
place within the nuclear medium. 
The dynamics of the nucleons within the nuclear matter are described through 
the relativistic mean field Lagrangian of Walecka \cite{4W,4SW}. 
The $\eta$ meson is described by solutions to the Klein-Gordon equation.
The interactions of the fields involved in the production reaction 
depicted in Fig.2, are described through the following interaction Lagrangian 
\begin{eqnarray}
{\cal L}_{\mbox{\small int}}={\cal L}_{\eta {\small N N}}
+{\cal L}_{\gamma {\small N N}}+{\cal L}_{V \eta\gamma}
+{\cal L}_{V NN}+{\cal L}_{\eta {\small N R}}
+{\cal L}_{\gamma {\small N R}}.
\end{eqnarray}
In the present work we adopt the pseudoscalar form for the $\eta$ meson-nucleon states coupling.
This choice is based on the work of Bennhold {\it et al.} \cite{4BEN.NPA} and 
Lee {\it et al.} \cite{4LEE.NPA} 
which showed preference for this form in the production on protons. 
The interaction Lagrangians which involve the nucleon and spin-$\frac{1}{2}$ 
resonances are cast in the following forms \cite{4BEN.PRD}: 
\begin{eqnarray}
{\cal L}_{\eta {\small N N}}= &-& i g_\eta \overline{\psi} \gamma_{5}
 \psi \eta
\label{lagrangian}
\end{eqnarray}
\begin{eqnarray} 
{\cal L}_{\gamma {\small N N}} = &-& e \overline{\psi}\gamma_\mu { A }^\mu
                                                    \psi 
                             - \frac{e \kappa_{p}  }{4 M} \overline{\psi}
                                      \sigma^{\mu \nu} \psi F_{\mu \nu},
 \end{eqnarray}
\begin{eqnarray} 
{\cal L}_{\small V N N}  =   &-&  g_v \overline\psi\gamma_\mu\psi V^\mu  
                              -\frac{g_t}{4M}\overline\psi\sigma_{\mu\nu}
                                                 \psi V^{\mu\nu}, 
 \end{eqnarray}
\begin{eqnarray} 
{\cal L}_{V \eta\gamma}=                     & &\frac{e\lambda_V}{4m_\eta}
                                 \epsilon_{\mu\nu\lambda\sigma}F^{\mu\nu}
                                                V^{\lambda\sigma}\eta , 
\end{eqnarray}
\begin{eqnarray} 
{\cal L}_{\eta {\small N R}}=  &-&ig_{\eta N R}\overline\psi\Gamma R\eta +h.c. 
   \label{s1} 
\end{eqnarray}
\begin{eqnarray} 
{\cal L}_{\gamma {\small N R}} = &-& \frac{e \kappa^R_{p} }{2(M_R+ M)} 
                      \overline{R}\Gamma^{\mu \nu}\psi F_{\mu \nu}+h.c. , 
\label{s2}
\end{eqnarray}
\begin{eqnarray} 
&&\hspace{-2 cm}\mbox{where} 
\nonumber \\
&&\hspace{-2 cm}\mbox{for $S_{11}$ resonances \hspace{1 cm} $\Gamma = 1$
~and~$\Gamma^{\mu\nu}=\gamma_5\sigma^{\mu \nu}$} 
\nonumber \\
&&\hspace{-2 cm}
\mbox{for $P_{11}$  resonances \hspace{1 cm} $\Gamma = \gamma_5$
~and~$\Gamma^{\mu\nu}=\sigma^{\mu \nu}$}
 \label{eq6}
\end{eqnarray}
$\kappa_p$ and $\kappa^R_p$ are the anomalous magnetic moment of the 
proton and the nucleon (~proton~) resonance, respectively and M$_R$ is the 
resonance mass.
The tensor $V_{\mu\nu}(x)$ is related to the vector meson field $V_\mu$ by 
\begin{eqnarray} 
 V_{\mu\nu}(x)=\partial_\mu V_\nu (x)-\partial_\nu V_\mu (x)
\label{eq5}
\end{eqnarray}  
The interaction of photons and $\eta$ mesons with the spin-$\frac{3}{2}$ 
( D$_{13}$(1520) )
resonance are introduced through the following Lagrangians,
\begin{eqnarray} 
{\cal L}_{\eta {\small N R}}=& &\frac{f_{\tiny \eta N R}}{m_\eta}\overline{R}^\mu
\theta_{\mu\nu}(Z)\gamma_5 \psi\partial^{\nu}\eta+h.c., 
\label{res3hf}           \\ 
{\cal L}^{1}_{\gamma {\small N R}}=  & &\frac{ie\kappa^{1}_R}{2M} 
\overline{R}^\mu\theta_{\mu\nu}(Y)\gamma_\lambda \psi F^{\nu\lambda} +h.c.,
           \\
{\cal L}^{2}_{\gamma {\small N R}}=  & &\frac{e\kappa^{2}_R}{4M^2}
 \overline{R}^\mu\theta_{\mu\nu}(X)\partial_\lambda N F^{\nu\lambda} +h.c.,
\label{eq6a}
\end{eqnarray}
where R$^\mu$ is the field associated with the D$_{13}$(1520) resonance. 
Two anomalous magnetic moments 
$\kappa^1_R$ and $\kappa^2_R$ are used in conjunction with the two different 
interaction Lagrangians introduced by Benmerrouche {\it et al.} \cite{4BEN.PRD}
for the $\gamma\eta R$ vertex.
The function $\theta_{\mu\nu}(V)$ is given by
\begin{eqnarray} 
 \theta_{\mu\nu}(V)&=&g_{\mu\nu}+\left[-\frac{1}{2}(1+4V)  +V\right]
\gamma_\mu\gamma_\nu,
\nonumber\\
 V&=&X,Y,Z
\label{eq7}
\end{eqnarray}  
where the parameters X, Y and Z, often referred to as off shell parameters, 
are obtained by fitting the experimental data for the elementary reaction. 

The authors of reference \cite{4BEN.PRD}, through fits to existing data, 
provided several sets of values for the parameters and observables related to 
the coupling constants of the effective interaction Lagrangians of 
Eqs. (\ref{lagrangian}-\ref{eq6a}). 
The parameters we use are those of the second column of table V of the above 
reference. 
These parameters are associated with the PS coupling which is the form used 
in the present work. 
The off-shell parameters $\alpha$, $\beta$ and $\delta$ given in that table 
are related to the X, Y and Z introduced in Eqs. (\ref{res3hf}-\ref{eq6a}), as
\begin{eqnarray}  
\alpha=1+4Z,\hspace {1.5 cm}\beta =1+4y,\hspace {1.5 cm}\delta=1+2X
\label{eq7a}
\end{eqnarray}  
A total of five nuclear resonances are included in the present calculations.
Table \ref{resonances} gives some of the properties of these resonances.

\subsection{The Reaction Amplitude}
 
At the tree-level the S-matrix for the A$(~\gamma,~\eta p~)$A-1 reaction 
is \cite{4MANSHA}
\begin{equation}
S_{fi} = -\frac{1}{2}  \int{<f|T\left[{\cal L}_{int}\left(x\right)
{\cal L}_{int}\left(y\right)\right]|i>  d^{4}xd^{4}y }
\label{eq8}
\end{equation}
where $|i>$ and $|f>$ are the initial and final states of the system, 
respectively.  
The T in front of the square bracket denotes the time ordered product of 
the operators within the bracket.
We shall illustrate, below, the derivation of the S-matrix for one of the 
contributing diagrams of Fig.2, namely the $S_{11}$ diagram. 
The same procedure can be used to derive the S-matrix for the other diagrams.

After introducing ${\cal L}_{\gamma NR} $ and ${\cal L}_{\eta NR} $ 
( from Eqs. (\ref{s1}), (\ref{s2}) and (\ref{eq6}) )
and using Wick's theorem \cite{4MANSHA}, 
the S-matrix for the S$_{11}$ resonance diagrams can be written as
\begin{eqnarray}
 \hspace{-.35 in}S^R_{fi} = -\frac{ie\kappa^R_p g_{\eta NR}}{2(M+M_R)}\int 
 <f|N\left[\bar{\psi}\left(y\right)\eta\left(y\right)
\underline{R\left(y\right)\bar{R}
\left( x\right)} \gamma_5\sigma_{\mu\nu}F^{\mu\nu}\left(x \right)
\psi\left( x\right)\right.
 \nonumber\\ 
 \hspace{.35 in}\left.+\bar{\psi}\left( x\right) \gamma_5
\sigma_{\mu\nu}F^{\mu\nu}\left(x \right)\underline{R\left( x\right)\bar{R}
\left(y\right)}\psi\left(y\right)\eta
\left(y\right)\right]|i> dx^4dy^4.
\label{eq11}
\end{eqnarray}
The initial state $|i>$ contains a photon and a target nucleus
with A nucleons. 
The latter is regarded as made up of a core and a valance nucleon. 
We write
\begin{eqnarray} 
|i> =  & &a^{\dagger}_{s_\gamma}\left({\bf k}_{\gamma}\right)
 \sum_{J_B M_B J}{\left( J, J_B; M, M_B| J_i, M_i \right) }
\nonumber \\  
       & &\times{ \left[ {\cal S}_{J_i J_f} (J_B) \right] }^{1/2}
         b^{\dagger}_{J_B M_B} |\phi_{J}^{M}>,                                   
\label{eq12}
\end{eqnarray}  
where $a^{\dagger}_{s_\gamma}\left({\bf k}_{\gamma}\right)$ is the 
creation operator for a photon with momentum ${\bf k}_{\gamma}$ and 
polarization $s_\gamma$. $b^{\dagger}_{J_B M_B}$ creates a nucleon 
with angular momentum quantum numbers
$J_B$ and $M_B$ bound to the core $|\phi_{J}^{M}>$ and ${\cal S}$ 
is the corresponding spectroscopic factor. 

The final state $| f >$, composed of an $\eta$ meson, a nucleon and the 
residual nucleus, is written in a similar fashion using creation operators 
for the $\eta$ and the nucleon.
To reduce the complexity of the numerical calculations we use the plane 
wave approximation for the propagators \cite{4JON.NPA1}. 
In this approximation the propagator for a spin-$\frac{1}{2}$ resonance 
assumes the simple form
\begin{eqnarray}
\underline{R(x)\bar{R}(y)} ~ =~ iS_R^F(x-y)~=~ i\frac{{ / \hspace{-0.1 in} 
k_R }+M_R}{k_R^2-M_R^2}\delta^4(x-y),
\label{eq20}
\end{eqnarray}
where $ S_R^F(x-y)$ is the Feynman propagator of a nucleon resonance with 
mass M$_R$ and four momentum $k_R$.
Carrying out the Fock space calculations and using the radiation gauge,
 the S-matrix of
 Eq. (\ref{eq11}) reduces to the following form
\begin{eqnarray}
S^R_{fi} &&= 
\frac{e}{ \left( 2\pi \right)^{9/2}}
{\left( \frac{M}{E_p} \frac{1}{2E_\eta} \frac{1}{2E_\gamma} \right)}^{1/2}
\nonumber\\  
&&\times \sum_{J_B M_B}{\left( J_f, J_B; M_f, M_B| J_i, M_i \right)
{ \left[ {\cal S}_{J_i J_f} (J_B) \right] }^{1/2}}
\nonumber\\  
&&\times \int{d^4x \Psi_{s_f} ^{(-)\dagger} (x)
\Gamma_{S_{11}} \Psi_{J_BM_B}(x)
\varphi_{\eta}^{\ast} (x)
e^{- i {  k_{\gamma} \cdot  x  } } },
\label{eq19}
\end{eqnarray}
where $\Gamma_{S_{11}}$ is a 4$\times$4 matrix operator and contains 
combinations of the Dirac $\gamma$ matrices, polarization of the interacting 
photon, mass and four momentum of the propagating resonance.  
The explicit form of this operator for each of the contributing diagrams are 
given below. In the above equation, E$_\gamma$, E$_p$ and E$_\eta$ are energies 
of the photon, proton and 
$\eta$ meson respectively and M is proton mass. $\Psi_{J_B,M_B}(x)$ describes 
a bound nucleon and is a solution of Dirac 
equation with the appropriate strong scalar and vector potentials. 
$\Psi_{s_f}(x)$ is the wavefunction of the outgoing proton which in the plane 
wave approximation is just a plane wave Dirac spinor.  
In the distorted wave calculations $\Psi_{s_f}(x)$ is a solution of the 
Dirac equation with the strong vector and scalar optical potentials. 
The $\eta$ meson wavefunction $\varphi_{\eta}$ is a solution of the Klein-Gordon equation with an appropriate optical potential for the 
$\eta$-nucleus system. 

The structure of the total S-matrix is the same as $S^R_{fi}$ of Eq. 
(\ref{eq19}) except for $\Gamma_{S_{11}}$ which must be replaced by a sum 
representing contributions from all the diagrams included in the model ( Fig.2 ). 
We obtain the following operators for these diagrams
\begin{eqnarray}
\Gamma_{proton}& =& g_{\eta {\small NN}}\gamma_0\left(\gamma_5
\frac{ { / \hspace{-0.095 in} k_{s} } + M }{ k^2_{s} - M^{2} } 
(\frac{\kappa_p}{2M}{ { / \hspace{-0.105 in} k }_{\gamma}+1) 
{ / \hspace{-0.1 in} \epsilon_r } }
\right. \nonumber \\
& & \hspace {1.5 cm} \left.
+(\frac{\kappa_p}{2M}{ { / \hspace{-0.105 in} k }_{\gamma}+1)
{ / \hspace{-0.1 in} \epsilon_r } }
\frac{ { / \hspace{-0.095 in} k_{u} } + M }{ k^2_{u} - M^{2} }
\gamma_5 \right), \label{gammp}
\end{eqnarray}
\begin{eqnarray}
\Gamma_{S_{11}} = \frac{g_{\eta {\small NR}}\kappa_p^R }{M+M_R} 
\gamma_0 \left(
\frac{ { / \hspace{-0.095 in} k_{s} } + M_R }{ k^2_{s} - M_R^{2} } 
\gamma_5
{ { / \hspace{-0.105 in} k }_{\gamma} { / \hspace{-0.1 in} 
\epsilon_r } }
+\gamma_5 { { / \hspace{-0.105 in} k }_{\gamma} 
{ / \hspace{-0.1 in} \epsilon_r } }
\frac{ { / \hspace{-0.095 in} k_{u} } + M_R }
{ k^2_{u} - M_R^{2} }\right),
\end{eqnarray}
\begin{eqnarray}
\Gamma_{P_{11}} = \frac{g_{\eta {\small NR}}\kappa_p^R }{M+M_R} 
\gamma_0 \left(\gamma_5
\frac{ { / \hspace{-0.095 in} k_{s} } + M_R }{ k^2_{s} - M_R^{2} }
{ { / \hspace{-0.105 in} k }_{\gamma} { / \hspace{-0.1 in} 
\epsilon_r } }
+ { { / \hspace{-0.105 in} k }_{\gamma} { / \hspace{-0.1 in}
 \epsilon_r } }
\frac{ { / \hspace{-0.095 in} k_{u} } + M_R }{ k^2_{u} - M_R^{2} }
\gamma_5\right),
\end{eqnarray}
\begin{eqnarray}
\Gamma_{V}^v =-i \frac{\lambda_v g_v}{m_\eta}
\epsilon_{\mu\nu\lambda\sigma } \gamma_0 
\frac{\epsilon^\mu k^\nu_\gamma k^\lambda_\eta\gamma^\sigma }
{k_t^2-m_V^2},
\end{eqnarray}
\begin{eqnarray}
\Gamma_{V}^t =\frac{\lambda_t g_t}{2Mm_\eta}
\epsilon_{\mu\nu\lambda\sigma} \gamma_0 
\frac{\epsilon^\mu k^\nu_\gamma\sigma^{\sigma\alpha}k_{t\alpha} k^\lambda_\eta}
{k_t^2-m_V^2},
\end{eqnarray}
\begin{eqnarray}
&&\hspace{-.2 in}\Gamma^{(1)}_{D_{13}}= -\frac{f_{\eta {\small NR}}
\kappa_R^{(1)} }{2Mm_\eta} \gamma_0 
\nonumber\\ \nonumber
&&\times\left(\gamma_5
k_\eta^\nu\theta_{\nu\mu}(Z)
\frac{ { / \hspace{-0.095 in} k_{s} } + M_R }{ k^2_{s} - M_R^{2} }
P^{\mu\alpha}\theta_{\alpha\beta}(Y)\gamma_{\lambda}
(k^\beta_\gamma\epsilon^\lambda_r-\epsilon^\beta_rk^\lambda_\gamma)\right.
\nonumber     \\ 
&&\hspace{.2 in}+\left.\gamma_{\lambda}\theta_{\beta\alpha}(Y)
(k^\beta_\gamma\epsilon^\lambda_r-\epsilon^\beta_r k^\lambda_\gamma)
\frac{ { / \hspace{-0.095 in} k_{u} } + M_R }{ k^2_{u} - M_R^{2} }
P^{\alpha\mu}\theta_{\mu\nu}(Z)k_\eta^\nu\gamma_5\right),
\end{eqnarray}
\begin{eqnarray}
&&\hspace{-.2 in}\Gamma^{(2)}_{D_{13}}= \frac{f_{\eta {\small NR}}
\kappa_R^{(2)} }{4m_\eta M^2} \gamma_0 
\nonumber\\ \nonumber
&&\times\left(\gamma_5
k_\eta^\nu\theta_{\nu\mu}(Z)
\frac{ { / \hspace{-0.095 in} k_{s} } + M_R }{ k^2_{s} - M_R^{2} }
P^{\mu\alpha}\theta_{\alpha\beta}(X)
(k^\beta_\gamma\epsilon^\lambda_r-\epsilon^\beta_rk^\lambda_\gamma)
\left\{ k^B \right\}_{\lambda}\right.
\nonumber\\ 
&&\hspace{.2 in}+\left.\left\{ k_p \right\}_{\lambda}
\theta_{\beta\alpha}(X)
(k^\beta_\gamma\epsilon^\lambda_r-\epsilon^\beta_r k^\lambda_\gamma)
\frac{ { / \hspace{-0.095 in} k_{u} } + M_R }{ k^2_{u} - M_R^{2} }
P^{\alpha\mu}\theta_{\mu\nu}(Z)k_\eta^\nu\gamma_5\right),
\label{eq21}
\end{eqnarray}
where
\begin{eqnarray}
\hspace{-.2 in}P^{\mu\nu} = \left(g^{\mu\nu}-\frac{1}{3}
\gamma^\mu\gamma^\nu -\frac{1}{2M_R}
\left[\gamma^\mu k^\nu-\gamma^\nu k^\mu  \right] 
-\frac{2}{3M_R^2}k^\mu k^\nu \right),
\label{eq21.1}
\end{eqnarray}
and $k^B$ is the local momentum of the bound nucleon and
\begin{eqnarray}
  k_s = k_\eta + k_p,\hspace{1.5 cm}
k_t = k_\gamma -k_\eta ,
\hspace{1 cm}\mbox{       }\hspace{1.2 cm}k_u = k_p - k_\gamma.
\label{eq22}
\end{eqnarray}

\subsubsection{Plane Wave Calculations}
To calculate the amplitude of Eq. (\ref{eq19}) one can use either 
plane waves or distorted waves to describe the outgoing particles.
In the plane wave approximation the final state interactions of the 
outgoing nucleon and $\eta$-meson, with the recoil nucleus, are ignored. 
The respective wavefunctions have the form
\begin{eqnarray}
\psi_p(k_p,x)&=&\sqrt{\frac{E+M}{2M}}\left (\matrix{
1\cr
\frac{\mbox{\boldmath{$\sigma$}}\cdot \mbox{\boldmath{$k$}}_p}{E+M}
\cr
}\right )e^{-ik_p\cdot x}\chi_{s_f},
\nonumber\\ 
\varphi_\eta(x)&=&e^{-ik_\eta \cdot x}.
\label{eq23}
\end{eqnarray}
Thus in the plane wave approximation (~PWA~) the S-matrix in Eq. (\ref{eq19}) 
takes the form
\begin{eqnarray}
S^R_{fi} &&= 
\frac{e}{ \left( 2\pi \right)^{7/2}}
{\left( \frac{M}{E_p} \frac{1}{2E_\eta} \frac{1}{2E_\gamma} \right)}^{1/2}
\delta(E_B+E_\gamma-E_p-E_\eta)
\nonumber\\  
&&\times \sum_{J_B M_B}{\left( J_f, J_B; M_f, M_B| J_i, M_i \right)
{ \left[ {\cal S}_{J_i J_f} (J_B) \right] }^{1/2}}
\nonumber\\  
&&\times \int{d^3x u^\dagger_{sf}(\mbox{\boldmath{$k$}}_p) 
e^{i\mbox{\boldmath{$k$}}_{rec}\cdot\mbox{\boldmath{$x$}}}
\Gamma_{S_{11}} \Psi_{J_BM_B}(\mbox{\boldmath{$x$}})},
\label{eq24}
\end{eqnarray}
where $\mbox{\boldmath{$k$}}_{rec} $ is the momentum of the recoil nucleus
 ( $\mbox{\boldmath{$k$}}_{rec} =\mbox{\boldmath{$k$}}_{\gamma}-
\mbox{\boldmath{$k$}}_{p}-\mbox{\boldmath{$k$}}_{\eta} $ ). 
Equation (\ref{eq24}) shows that the PWA S-matrix is made up of terms 
proportional to the Fourier transform of the components of the bound state 
wavefunction.

\subsubsection{Final State Interactions: Distorted Wave Calculations}
The rest of this section is devoted to a derivation of the form of the S-matrix in 
the distorted wave approximation (~DWA~).
The continuum nucleon wavefunction is written as
\begin{eqnarray}
\psi_{s_f}^{(-)^\dagger}(x) &=& 4\pi \left[\frac{E+M}{2M}\right]^{1/2} e^{iEt}
\sum_{LJM} i^{-L} Y^{M-s_f}_{L}(\hat{\bf  k}_f)  
                                  {\cal Y}^{M^\dagger}_{L1/2J}(\Omega) 
\nonumber\\ 
& & \times ( L,1/2; M-s_f, s_f| J, M )
\left [ \matrix{ f_{LJ} (r), & i\sigma\cdot\hat{\bf r}g_{LJ}(r) \cr } \right ]
\label{eq27}
\end{eqnarray}
where 
f$_{LJ}$ and g$_{LJ}$ refer to the upper and lower radial functions. 
These are solutions of the radial Dirac equation with 
vector and scalar optical potentials obtained from comparison with 
proton-nucleus elastic scattering data. 
More details of the Dirac proton wavefunctions can be found in reference 
\cite{4JON.NPA1}.
Expanding the eta and photon wavefunctions into partial waves and using 
these in the  expression of the amplitude given by Eq. (\ref{eq19}), 
one gets the DWA S-matrix as
\begin{eqnarray}
S_{fi} &=& \frac{ e }{\pi}
{\left( \frac{E_p + M}{E_p E_\eta E_\gamma} \right)}^{1/2}
\delta( E_B + E_\gamma - E_p  - E_\eta ) 
\nonumber \\
& &\times \sum_{ J_B M_B}{ \left( J_f, J_B; M_f, M_B| J_i, M_i \right)
                     { \left[ {\cal S}_{J_i J_f} (J_B) \right] }^{1/2}  } 
\nonumber \\
& &\times \sum_{L J L_\eta L_\gamma}
                       { i^{L_\gamma - L - L_\eta } 
\left( 2 L_\gamma + 1 \right)^{1/2} } 
 \nonumber\\
& &\times \sum_{ M M_\eta}{ Y_L^{M - s_f}\left( { \bf \hat{k}}_p  \right)
   \left[Y_{L_\eta}^{M_\eta}\left( { \bf \hat{k}_\eta } \right) \right]^{\ast}    
           \left( L, \frac{1}{2}; M - s_f, s_f| J, M \right)           }  
\nonumber \\
& &\times
\begin{array}[t]{l} 
\left\{
\ R_{ff} \left[ \kappa \kappa_B L_\eta L_\gamma \right]
  A_{11}{\left[ \matrix{ \kappa \cr M \cr } 
\matrix{ \kappa_{B} \cr M_{B} \cr }
\matrix{ L_{\eta} \cr M_{\eta} \cr } L_{\gamma} \ \xi\right] } \right. 
\nonumber \\
 \\
+ i R_{fg} \left[ \kappa \kappa_B L_\eta L_\gamma \right]
A_{12}{\left[ \matrix{ \kappa \cr M \cr } \matrix{ \kappa_{B}^\prime 
\cr M_{B} \cr }
         \matrix{ L_{\eta} \cr M_{\eta} \cr } L_{\gamma} \ \xi      
            \right] } \\
 \\
- i R_{gf} \left[ \kappa \kappa_B L_\eta L_\gamma \right]
A_{21}{\left[ \matrix{ \kappa^\prime \cr M \cr } \matrix{ \kappa_{B} 
\cr M_{B} \cr }
         \matrix{ L_{\eta} \cr M_{\eta} \cr } L_{\gamma} \ \xi         
         \right] } \\
 \\
\left.+ \ R_{gg} \left[ \kappa \kappa_B L_\eta L_\gamma \right]
             A_{22}{\left[ \matrix{ \kappa^\prime \cr M \cr }
                  \matrix{ \kappa_{B}^\prime \cr M_{B} \cr }
       \matrix{ L_{\eta} \cr M_{\eta} \cr } L_{\gamma} \ \xi   
       \right]  }  \right\}
\label{eq29}
\end{array}    
\\
\end{eqnarray}
where $\kappa = ( L - J ) ( 2J + 1 )$ and $L^\prime = 2J - L$. 
The radial integrals are given by
\begin{eqnarray}
R_{CB} \left[ \kappa \kappa_B L_\eta L_\gamma \right] = 
\int{C_{LJ}\left(r\right)B_{L_{B}M_{B}} \left(r\right
)v_{L_\eta}\left(k_\eta r\right) j_{L_\gamma}\left(k_\gamma r\right) r^2 dr},
\end{eqnarray}
and $C\left(r\right)$ is either an upper- or lower-component radial 
function for the ejected proton while $B\left(r\right)$ 
is the counterpart for the bound state wavefunction. The angular 
integrals are given as follows
\begin{eqnarray}
A_{ij}{\left[ \matrix{ \kappa^\prime \cr M \cr } 
\matrix{ \kappa_{B} \cr M_{B} \cr }
\matrix{ L_{\eta} \cr M_{\eta} \cr } L_{\gamma} \ \xi  \right] }
=
\int {\left({\cal Y}^M_{L1/2J}\right)^\dagger\Gamma_{ij}
{\cal Y}^{M_B}_{L_B1/2J_B}Y^{M_\eta}_{L_\eta}
\left( \Omega\right)Y^{0}_{L_\gamma}
\left( \Omega\right)d\Omega}, 
\label{eq30}
\end{eqnarray}		      
The matrix operators $\Gamma_{ij}$ are the elements of the 4$\times$4 
$\Gamma$ matrix 
operator 
\begin{eqnarray}
\Gamma=\pmatrix{
\Gamma_{11} & \Gamma_{12} \cr
\Gamma_{21} & \Gamma_{22} \cr}.
\label{eq31}
\end{eqnarray}

\subsection{Observables}
It is useful to define a function of the initial and final spin projections 
$Z^{s_f}_{\xi M_B}$ by rewriting the amplitude of Eq. (\ref{eq29}) 
in the following form 
\begin{eqnarray}
S_{fi} &=& \frac{ e }{\pi} 
{\left( \frac{E_p + M}{E_p E_\eta E_\gamma} \right)}^{1/2}
\delta( E_p + E_\gamma - E_B - E_\eta ) 
\nonumber \\
& &\times \sum_{ J_B M_B}{ \left( J_f, J_B; M_f, M_B| J_i, M_i \right)}
\nonumber \\ 
& &\hspace{.1 in} \times { \left[ {\cal S}_{J_i J_f} (J_B) \right] }^{1/2} 
                                                     Z^{s_f}_{\xi M_B}
\label{eq32}
\end{eqnarray}
The triple differential cross section is related to $Z^{s_f}_{\xi M_B}$ by
\begin{eqnarray}
\frac{d^3\sigma}{d\Omega_{\eta} d\Omega_p dE_\eta}  &=& 
       \frac{8\alpha \pi}{\hbar c}
            {\left[ \frac{E_p + M c^2} {E_{\gamma} } \right]} 
            k_p c \; k_{\eta} c
\nonumber\\  
& &\times\frac{1}{R} \sum_{J_B M_B s_f \xi }{ \frac{ {\cal S}_{J_i J_f} (J_B) }
{ 2J_B + 1 } }{ | Z^{s_f}_{\xi M_B} | }^2
\nonumber\\                   
\label{eq33}
\end{eqnarray}
where $\alpha$ is the fine structure constant and the recoil factor R is defined 
in a manner similar to the case of ( $e,~e'p$ ) reactions \cite{4RECOIL}
\begin{eqnarray}
R=1-\frac{E_p}{E_r}\frac{1}{\mid {\bf k}_p\mid^2}{\bf k}_p\cdot{\bf k}_r
\nonumber\\  
\label{eq34}
\end{eqnarray}
and E$_r$ is the total energy of the residual nucleus.

The photon asymmetry for linearly polarized incident photons is
\begin{eqnarray}
A = \frac{ d \sigma_{\parallel} - d \sigma_{\perp}  } 
{ d \sigma_{\parallel} + d \sigma_{\perp} }
\label{eq35}
\end{eqnarray}
where d$\sigma_{\parallel}$ and d$\sigma_{\perp}$ are the cross 
section for specified polarizations of the incident photon, 
namely parallel and perpendicular to the plane of 
the reaction.

\section{Results and Discussion }
This paper is concerned with the study of the exclusive quasifree 
(~$\gamma $, $\eta$p~) reaction on complex nuclei. 
To date there are no available experimental data for this particular 
reaction ( data exist only for inclusive reactions \cite{4pri} ). 
Since the lack of attempts to measure the exclusive observables is 
mainly due to the expectation of low cross sections, 
it is of practical importance to identify the kinematical 
regions in which the cross sections are relatively large.
Figure 3 shows the calculated cross section for this 
reaction on $^{12}$C leading to the ground state of $^{11}$B ( assumed to 
be a pure p$_{3/2}$ hole state ). 
The incident photon energy is 750 MeV.  
The calculations are carried out in the plane wave approximation for the 
outgoing proton and $\eta$ meson. The bound nucleon is 
described by a solution of the
Dirac equation involving the relativistic scalar and vector Hartree 
potentials of reference \cite{4HOR.NPA}. 
We used the maximum possible value for the spectroscopic 
factor (~i.e 2J$_B$ + 1; J$_B$ is the total angular momentum of the 
bound proton~).
The azimuthal angles of the outgoing proton and $\eta$ meson are fixed at 
180$^o$ and 0$^o$, respectively. 
Calculations are performed at four different polar angles for  
the $\eta$ meson and cover the outgoing proton polar angular range 
10$^o$-50$^o$. 
Cross sections are plotted as a function of polar angle of the outgoing proton 
and the kinetic energy of the $\eta$ meson.
These results indicate that one can observe relatively large cross sections 
for $\theta_\eta$ in the range 10$^o$-30$^o$ and $\theta_p$=15$^o$-30$^o$.
The pair of angles $\theta_\eta$ = $\theta_p$ = 30$^o$ may be a preferable 
setting experimentally and hence will be used for most of the analysis done here.
Similar results were found for $^{16}$O and $^{40}$Ca nuclei. 
These results are expected to change somewhat with respect to variation in the incident 
photon energy. 
The larger cross sections will shift towards more forward angles as the photon 
energy increases.

It is well known that the eta photoproduction reaction on the proton near 
threshold is dominated by the S$_{11}$(1535) resonance formation. 
We have done calculations aimed at assessing the extent to which this 
dominance prevails when the reaction takes place within the nuclear medium. 
The calculations are done for the reaction discussed above, again in the plane 
wave limit, where the angles of the outgoing particles are fixed at 
$\theta_\eta$ = $\theta_p$ = 30$^o$.
We varied the energy of the incident photon from near threshold to the 
energy where the contributions from the S$_{11}$(1535) diagrams are comparable 
to contributions from all other diagrams. 
Results of these calculations are illustrated in Fig.4: 
the dotted curves give the cross sections due to the formation 
of the S$_{11}$(1535); contributions 
due to the rest of the diagrams are represented by solid curves. 
Note the varying scale for the cross section axes. 
Figure 4 shows that the S$_{11}$(1535) resonance dominates 
the cross section from threshold up to a photon energy of about 1.1 GeV. 
The cross section is small at energies near threshold and increases with 
increasing photon energy, reaching its largest values in the photon energy 
range 750-950 MeV. 
At energies higher than 950 MeV, the cross section due to the S$_{11}$(1535) 
resonance decreases as the photon energy increases.
 
In Fig.5 we show the relative contributions of the individual diagrams.
The cross section arising from the S$_{11}$(1535) pole diagrams, 
shown by the long-dashed curve, is clearly the dominant contribution. 
The second largest contribution comes form the proton poles( 
double dash-double dotted curve ). 
The results of the calculations for the D$_{13}$(1520) resonance and the 
vector mesons poles are shown by dot-dot-dashed and dot-dash-dashed curves, 
respectively. 
The solid curve shows the total cross section ( i.e contributions from all 
the diagrams ). 
Besides confirming clearly that the S$_{11}$(1535) resonance is the main 
contributor to the reaction in this energy region, these results also give us 
an indication of the relative importance of other diagrams. 
The next largest contribution comes from the proton pole followed by the 
D$_{13}$ resonance, then the vector meson diagrams.
The P$_{11}$(1440), S$_{11}$(1650) and P$_{11}$(1710) resonances do not make 
significant contributions to the reaction at this energy.

In order to establish that the above conclusions are not effected by the 
inclusion of the final state interactions of the outgoing particles with the 
residual nucleus, we carried out distorted wave calculations in which 
these final state interactions are taken into account. 
Appropriate optical potentials are used in the calculation of the 
wavefunctions of the outgoing proton and $\eta$ meson. 
In all the distorted wave calculations presented here the global optical 
potentials of Cooper {\it et.al} \cite{4COO.PRC} have been used to calculate 
the Dirac distorted wavefunctions of the outgoing proton. 
For the outgoing $\eta$ meson we use the Klein-Gordon equation together 
with the optical potentials of references \cite{4LEE.NPA,4OSE.PRC}.
The results are shown in Fig.6; 
the solid curve shows the cross section due to the S$_{11}$(1535) resonance, 
whereas the dashed curve includes all contributions other than those of the 
S$_{11}$(1535).
These results indicate that, as in the plane wave calculations discussed above, 
the S$_{11}$(1535) resonance is the dominant contributor to the 
reaction in the distorted wave limit. 
Note the general suppression of the cross section as a result of the 
inclusion of the final state interactions.

Figure 7 shows the calculated observables for the reaction where we investigate 
separately the effects of the proton and $\eta$ meson final state interactions. 
The energy dependent global optical potential of Cooper {\it et al.} 
\cite{4COO.PRC}, 
and the optical potential DW1 of reference \cite{4LEE.NPA} are used for 
outgoing proton and $\eta$ meson, respectively. 
Plane wave and various distorted wave calculations are shown. 
The comparisons of the cross sections are shown in Fig. 7-(a) and those 
for the photon asymmetry are shown in Fig. 7-(b).
The short dashed curve shows the suppression due to the final state 
interaction of the outgoing proton with the residual nucleus. 
This suppression is more evident at higher $\eta$ energies.
These correspond to smaller energies of the outgoing protons and 
indicate that these protons are more affected by the distortion. 
This same qualitative behavior holds for the $\eta$ meson distortion 
(~dashed curve~); the suppression is skewed towards smaller $\eta$ 
energies. 
The dotted curve is the calculated cross section when distortion 
effects are included for both of the outgoing particles. 
The suppression of the cross section in this case is much more pronounced.

Figure 7-(b) shows the photon asymmetry for the same reaction
( see Eq. (\ref{eq35} )). 
Plane wave calculations (solid curve) produce a nearly flat curve for 
the photon asymmetry. 
Distortion of the outgoing $\eta$ meson affects the shape and magnitude 
of the photon asymmetry only slightly. 
Inclusion of the final state interactions of the outgoing proton 
results in larger changes in the asymmetry. 
When both final state interactions are included, the resulting curve 
(~dotted curve~) has characteristics close to the case when only 
the proton is distorted. 
 
Figure 8 shows the effects of different $\eta$ optical potentials on 
the calculated observables.  The curves of Fig.8 are obtained with the
 same bound and
continuum proton potentials while different optical potentials are used 
for the outgoing $\eta$ meson. Five different potentials are used. The first
 two are those (denoted by DW1 and DW2) used by  Lee {\it et al.} 
\cite{4LEE.NPA}.These 
are obtained in a $t\rho$ approximation using $\eta N$ t-matrices obtained in a 
coupled channel approach \cite{4BEN.NPA,4TIA.NPA,BSSN} 
( a recent work by Batinic  {\it et al.} \cite{BDSS} points out to a sign error 
in their analysis in \cite{BSSN} which may mean that 
the potential DW2 is not reliable ). 
The other sets of potentials are those of Chiang  {\it et al.} \cite{4OSE.PRC}. 
These are 
obtained by calculating the $\eta$ self energy in nuclear matter taking 
into account only contributions from the $S_{11}$ resonance. The three 
different potentials used correspond to three different choices of 
the real part of the $N^\ast$ self energy ( namely 
Re$\Sigma_{N^\ast}=(\rho /\rho_0)V_{N^\ast}$ with $V_{N^\ast}$=50, 0, -50 MeV; 
$\rho$ and $\rho_0$ are the nuclear density for finite and infinite nuclear 
matter, respectively. ) The uncertainty about 
the parameter $V_{N^\ast}$ is a drawback for the potentials of 
Chiang {\it et al.}.

In general the calculated cross sections show some dependence on the type 
of distorting potential used; 
there are some variations in shape and magnitude. 
The magnitude of the cross section obtained with the DW1 potential is 
somewhat larger than those predicted by the other potentials near 
the mid-range of $\eta$ energies.

The corresponding calculations for photon asymmetry with different optical 
potentials for the outgoing $\eta$ meson are shown in Fig.8-(b). 
All potentials produce approximately the same shape and 
magnitude for the photon asymmetry.
The asymmetry therefore, at least at this energy, is not particularly 
sensitive to the different choices of the final state interactions of 
$\eta$ mesons.

We have also calculated the observables of the reaction using several 
different proton optical potentials. 
We found that the calculated observables are rather insensitive to the 
variations among these optical potentials.

The sensitivity of the results to the bound state potentials was also 
assessed by performing calculations in which different binding potentials 
are used.
In this case relativistic scalar and vector potentials of Woods Saxon shape 
\cite{4GER.THS} were used to generate the Dirac bound state wavefunction. 
The results are shown by the dashed curve in Fig.9-(a) 
along with the results obtained using the Hartree potentials 
adopted throughout the present study (~solid curve~).  
Comparison of these curves indicates the level of sensitivity in our 
model calculations to the bound state potentials. 
This comparison shows that the Hartree bound state potentials lead to 
somewhat larger cross sections at this photon energy.
On the other hand the photon asymmetry calculations appear to be 
insensitive to different choices of the bound state potentials (Fig.9-(b)). 

In addition to the kinematics used above, we also calculated the 
observables of the reaction for another set of angles for the outgoing 
particles. The motivation for choosing this new set is to compare the 
results of our model with those of the nonrelativistic model of 
Lee {\it et al.} \cite{4LEE.NPA}. 
Figure 10 shows this comparison. The calculations are 
performed at the following angular settings for the outgoing particles: 
$\theta_\eta=20^o, ~\phi_\eta=0^o$ and $\theta_p=15^o, ~\phi_p=180^o$.
We have plotted the results of the non-relativistic plane wave 
calculations of Lee {\it et al.} \cite{4LEE.NPA} together with our results 
for both plane wave and distorted wave calculations. 
Figure 10-(a) shows the cross section and 10-(b) the resulting photon 
asymmetry.
Results of the relativistic plane wave and distorted wave calculations 
are shown by solid and short dashed curves, respectively.
The nonrelativistic plane wave calculations are depicted by the long dashed 
curves. 

Comparison of the solid and long dashed curves indicates that 
both relativistic and nonrelativistic calculations produce curves with 
similar shape, 
but the nonrelativistic calculations predict generally larger cross sections 
in the $\eta$ energy range up to $\sim$120 MeV, and fall below at higher $\eta$ 
energies. 
There is also a slight shift in the relative positions of the peaks 
in the two calculations.
As noted earlier, the inclusion of final state interactions leads to a 
suppression 
of the calculated cross sections.
 
The plane wave relativistic and nonrelativistic predictions for the photon 
asymmetry differ significantly 
( Note that Lee {\it et al.} use a definition of asymmetry which is opposite 
in sign to the one used here; 
for the purpose of the present comparison the results from reference 
\cite{4LEE.NPA} have been multiplied by -1 ).  
The relativistic calculations produce a flat curve with small positive values, 
whereas the nonrelativistic calculations result in a curve with large negative 
values, with strong dependence on the $\eta$ energy. 
The relativistic calculations show some sensitivity to final state 
interactions whereas Lee {\it et al.} report such effects are negligible 
in the nonrelativistic calculations. 

We carried out similar calculations for the photoproduction reaction on 
a $^{16}$O target leading to the ground state in $^{15}$N. 
Figure 11 shows the results of these calculations. 
The residual nuclear state is assumed to be a pure 1p$_{1/2}$ hole state.
The Hartree potentials of reference \cite{4HOR.NPA} are used to calculate 
the bound state 
wavefunction. 
The final state interactions of the outgoing proton and $\eta$ meson are 
calculated using 
the energy dependent global optical 
potential of reference \cite{4COO.PRC} and the DW1 optical potential 
of reference \cite{4LEE.NPA}, respectively. 

The results for the cross section are very much similar to the $^{12}$C case 
(~see Fig.7~). In the case of $^{16}$O we note that the distorted 
wave cross section 
( dotted curve ) is slightly smaller in comparison with the 
corresponding case in $^{12}$C 
( dotted curve in Fig. 7 ).
On the other hand the resulting photon asymmetries for this target, 
behave differently 
form those of the $^{12}$C target. 
The distorted wave calculations here peak at lower $\eta$ energies. 
Moreover asymmetries for this target 
have larger magnitudes but this occurs in a region of very small 
cross sections.
These differences with the $^{12}$C case indicate that the
asymmetry is dependent on the single particle state from which the 
proton is ejected.

We have also used our relativistic model for calculations of the reaction on 
$^{40}$Ca.
Figure 12 shows the results of our calculations for the 
case in which the residual nucleus ( $^{39}$K ) is left in its ground state 
(~a 1d$_{3/2}$ hole state~) 
and the incident photon energy is 750 MeV. 
The calculated cross sections of Fig.12-(a) show curves with 
different shapes from those of the two targets studied previously. 
The final state interactions of the outgoing particles lead to a suppression 
of the cross section for this target in a manner similar to what has been noted 
earlier for $^{12}$C and $^{16}$O. 
We find that the earlier observation on photon asymmetries is confirmed 
in the present case; the photon asymmetry 
calculations are sensitive to the final state interactions of 
the outgoing particles. 
This is in contrast to the predictions of the nonrelativistic 
model for the reaction \cite{4LEE.NPA}. 
The shape and magnitude of the asymmetries 
are different from those found for the  $^{12}$C 
and $^{16}$O targets. 

We have also carried out calculations for the 
observables of the reaction on a $^{40}$Ca target 
leading to an odd parity 1p$_{3/2}$ hole state in $^{39}$K. 
The results are shown in Fig.13.
The curves in this figure have the same features as in Fig.10, 
except that the differences in the magnitude of the relativistic and 
nonrelativistic 
cross sections are larger. 
Also the effects of the final state interactions are more prominent for 
the present case. 
Note that the photon asymmetries given in Figs. 10-(b) 
and 13-(b) have approximately the same shapes. These similarities and 
the fact that reaction is taking place on a proton in a 1p$_{3/2}$ 
state, support our earlier statement on the state dependency of the 
asymmetry.

The calculations reported in this section have all used a specific set 
of coupling constants; one of several that are given in reference 
\cite{4BEN.PRD}. 
The question then arises concerning the sensitivity of the calculated results to the possible differences among these sets which provide nearly equivalent fits to the data for the elementary reactions. 
We have carried out calculations using the set given in the first column of table V of reference \cite{4BEN.PRD}. 
The results show that there are some changes in the magnitudes of the calculated cross sections and photon asymmetries. These changes are at the level of 15\% for the cross sections but are somewhat larger for the asymmetries.

\section{Conclusions}

In this paper we have developed a relativistic model for the quasifree 
photoproduction of $\eta$ mesons on complex nuclei. 
The ingredients of the  model are: 
i) the nucleon wave functions are solutions of the Dirac equation with 
appropriate scalar and vector potentials consistent with the relativistic 
mean field approach,
ii) the $\eta$ meson is described by solutions of the Klein-Gordon equation 
with appropriate optical potentials, and 
iii) the interactions between the fields are introduced through a covariant 
effective Lagrangian.
Contributions from nucleon resonances, t-channel vector meson as well 
as the nucleon Born diagrams are included.
The model is used to calculate the cross section and photon asymmetry, 
for different target nuclei. Both plane wave and distorted wave calculations 
are presented.

We wish to comment on the question of the gauge invariance of the reaction 
amplitude calculated in the above manner. The effective Lagrangian we
start with is gauge invariant. As we look at the nuclear amplitude itself, 
as given by Eq. (\ref{eq19}), 
we find that its invariance is determined by the behavior of the 
$\Gamma$ operators given in Eqs.(\ref{gammp}-\ref{eq21}).  
All these operators are gauge invariant, except for the electric parts of 
Eq.(\ref{gammp}). 
These latter terms are invariant only in the limit where the particles involved in 
the elementary interaction are considered to be on their mass shell.

The energy region where the S$_{11}$(1535) resonance dominates 
the reaction is identified.
As expected this region is close to threshold and covers a range of 
photon energies from 750 MeV to 950 MeV.  
None of the other diagrams makes contributions at the same level as 
the S$_{11}$.
We find that the next leading diagrams are proton, D$_{13}$ 
resonance and vector meson poles. 
However we also find that there is some cancellation among these
contributions. 
In our view the energy region identified above is perhaps the best 
suited for using the $\eta$ photoproduction reactions to study the 
properties of the S$_{11}$ resonance in the nuclear medium. 

When the final state interactions of the outgoing particles are included, 
we find that the resulting cross sections are strongly suppressed. 
The extent of this suppression depends somewhat on the type of optical potential 
used to represent the final state interaction.
We find some variations among the available sets of potentials particularly 
those that describe the $\eta$ meson final state interactions. 
Several $\eta$ optical potentials were tested and lead to different results for the 
cross section as well as the photon asymmetry . 
The uncertainty in $\eta$ optical potentials makes it difficult to make final 
predictions for the cross sections. 
More theoretical work on the final state interactions of the $\eta$ is required.

Calculations were carried out using the same kinematics as those in the recent 
nonrelativistic 
calculations of Lee {\it et.al} \cite{4LEE.NPA}. 
Results of our calculations for the cross section have shapes close to  
those of these nonrelativistic calculations but the 
magnitudes of our results are somewhat smaller.
In contrast, the photon asymmetries predicted by the two models differ significantly. 
The nonrelativistic 
model gives large asymmetries in the plane wave limit, which are found to be 
insensitive to the final
state interactions of the outgoing particles. 
Our predictions yield much smaller asymmetries and these are 
strongly affected by the final state interactions.
Lee {\it et al.} have also found that a small percentage change in the masses 
of the $S_{11}$ and $D_{13}$ have large effects on the cross sections and 
the asymmetries, respectively. 
Our present calculations show similar effects. 
These are indications that eta photoproduction reactions can be used as 
a probe of medium modifications of the properties of these resonances. 
These of course are expected to go beyond the simple mass shifts; 
for example there may in fact be a radial density dependence of the 
masses of the propagators. Studies of these effects with the present 
relativistic model are planned.

The above statements indicate that the present reaction is rich in 
its physics content. 
If enough data become available, one will be able to clarify the role 
played by the intermediate nucleon states ( the $S_{11}$ and 
the $D_{13}$ ) as well as the vector mesons while embedded in the 
nuclear environment. 
The question of the optical potential 
for the $\eta$-nucleus system could also be clarified. 
The data could also be used to assess the need for a strictly 
relativistic treatment.
It is clear that there is a multiplicity of effects at play and hence 
it is imperative that measurements at several energies in the region 
discussed above and for a variety of target nuclei be carried out. 
Measurements of the cross sections and photon asymmetries in exclusive 
production, as well as coherent production,  
will be very helpful in shedding more light on the above issues.
In a subsequent paper the present model will be extended to the treatment of 
inclusive reactions for which some data have become available recently \cite{4pri}. 
\section*{Acknowledgements}
We are grateful to \H{J}. Ahrens, M. Benmerrouche, C. Bennhold, J.I. Johansson, 
B. Krusche, F.X. Lee, B. Lopez Alvaredo, 
E. Oset, M. Roebig-Landau, M. Vineyard and L.E. Wright for helpful 
communications concerning their work. 
M. Hedayati-Poor would like to acknowledge the support from The Iranian Ministry of 
Higher Education.

\newpage

\section* {Table Caption}

\noindent TABLE 1. Summery of the properties of the baryon resonances included in 
present model. 
J$^\pi$ is the spin parity, L is the relative angular momentum related to  
$\pi$N scattering state. The subscripts
I and J  for the L values (~third column~) represent the isospin and total angular 
momentum assigned to the resonance, respectively. $\Gamma$ is the total width of 
the resonance.
\vspace{.5 cm}

\section* {Figure Captions}

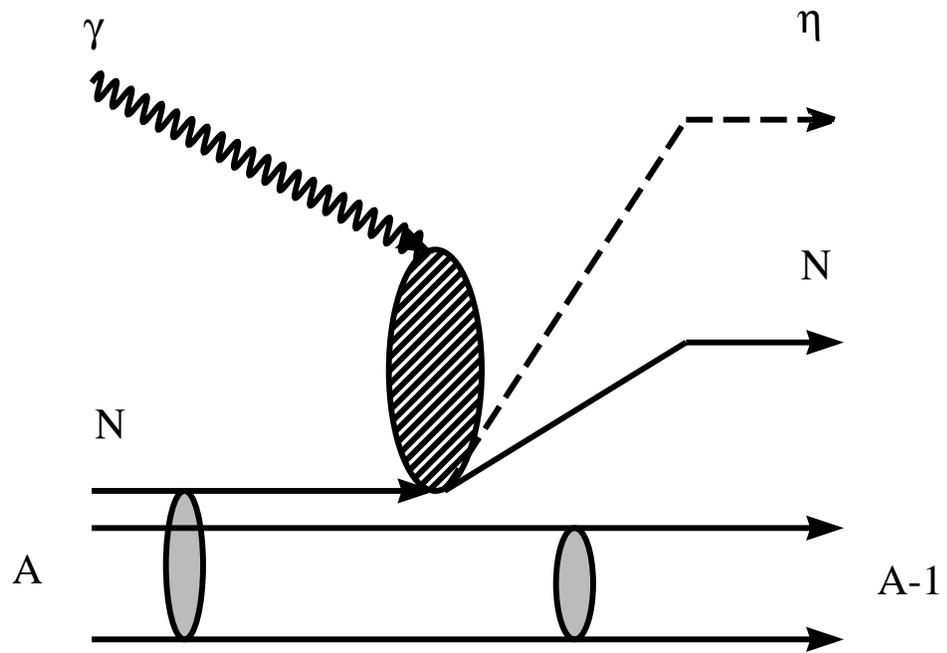
\noindent FIG. 1. Feynman diagram for an $A(~\gamma,\eta p~)A-1$ reaction. 
The incident photon with momentum ${\bf k}$ is absorbed by a 
proton embedded in nucleus A, resulting in ejection of a proton and an $\eta$ 
meson. 

\vspace{2.5 mm}
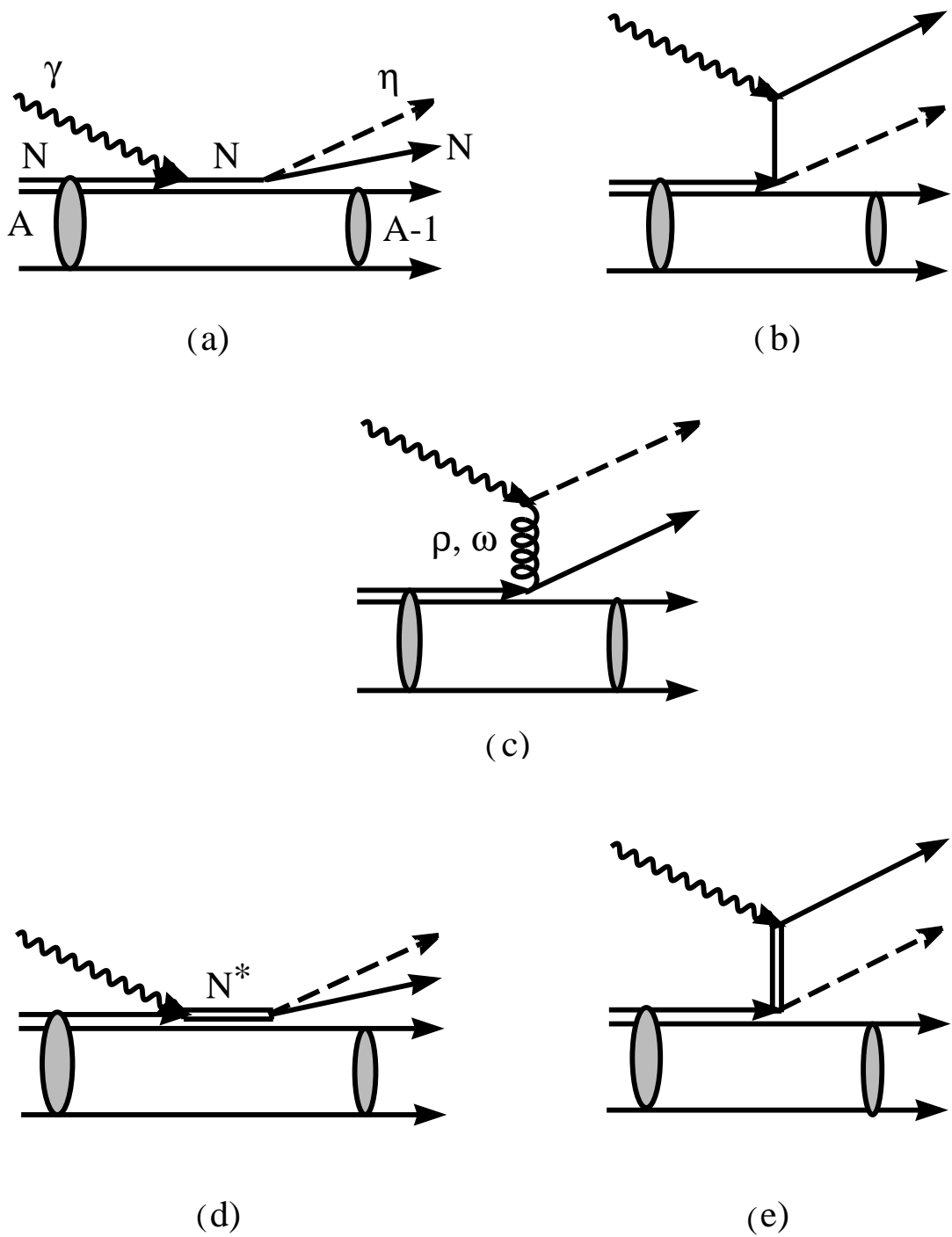
\noindent FIG. 2. The contributing Feynman diagrams to $A(~\gamma, \eta p~)A-1$ reaction.
 a ) and b) are the s- and u-channel Born diagram respectively, c) the t-channel 
vector meson diagram, d) and e) are the s- and u-channel nucleon resonance poles.  

\vspace{2.5 mm}
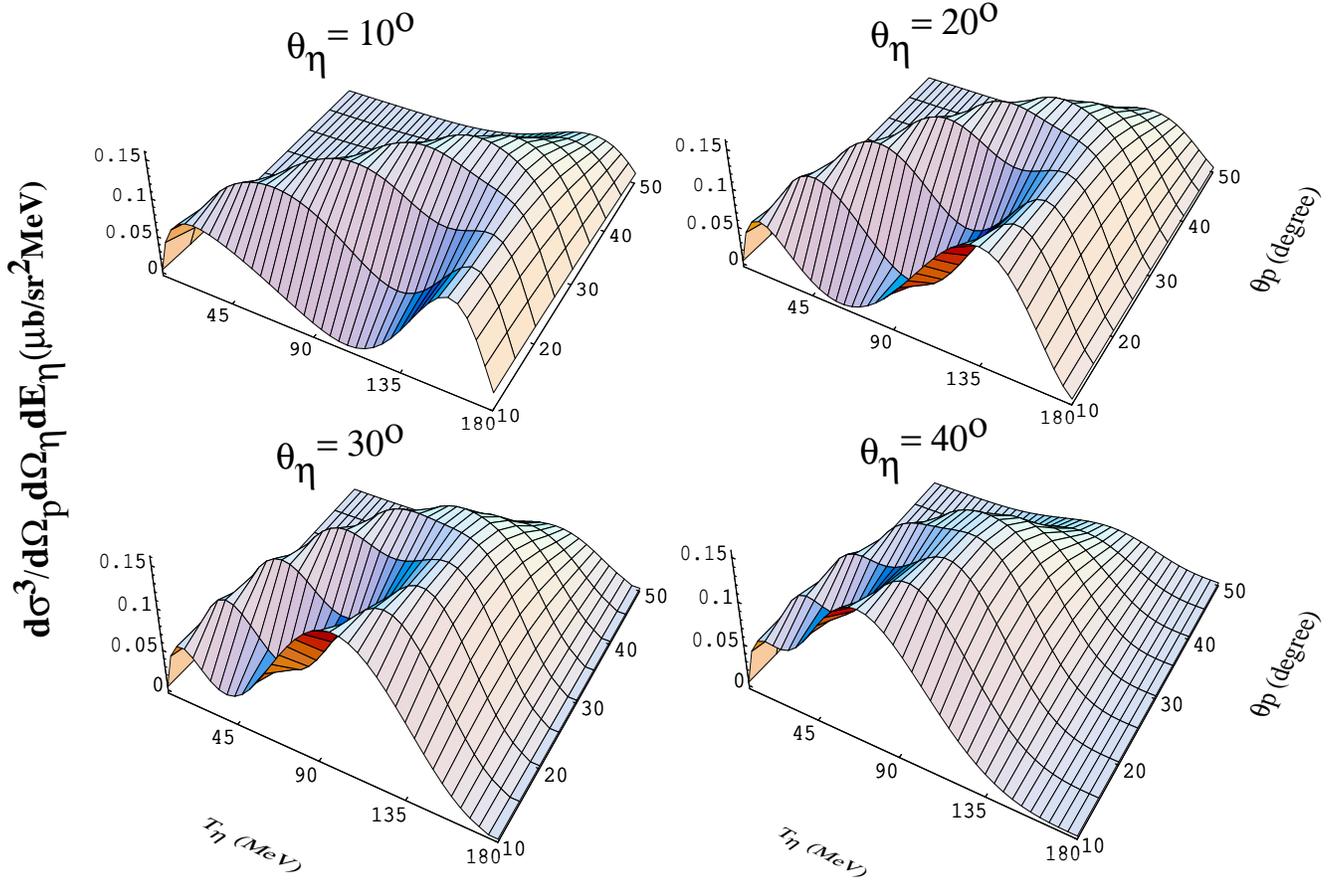
\noindent FIG. 3. The cross section for the $^{12}C(~\gamma, \eta p~)^{11}B_{g.s.}$ 
reaction for the regions of the phase space where the reaction has 
significant yield. Calculations are performed in the plane wave approximation.  

\vspace{2.5 mm}
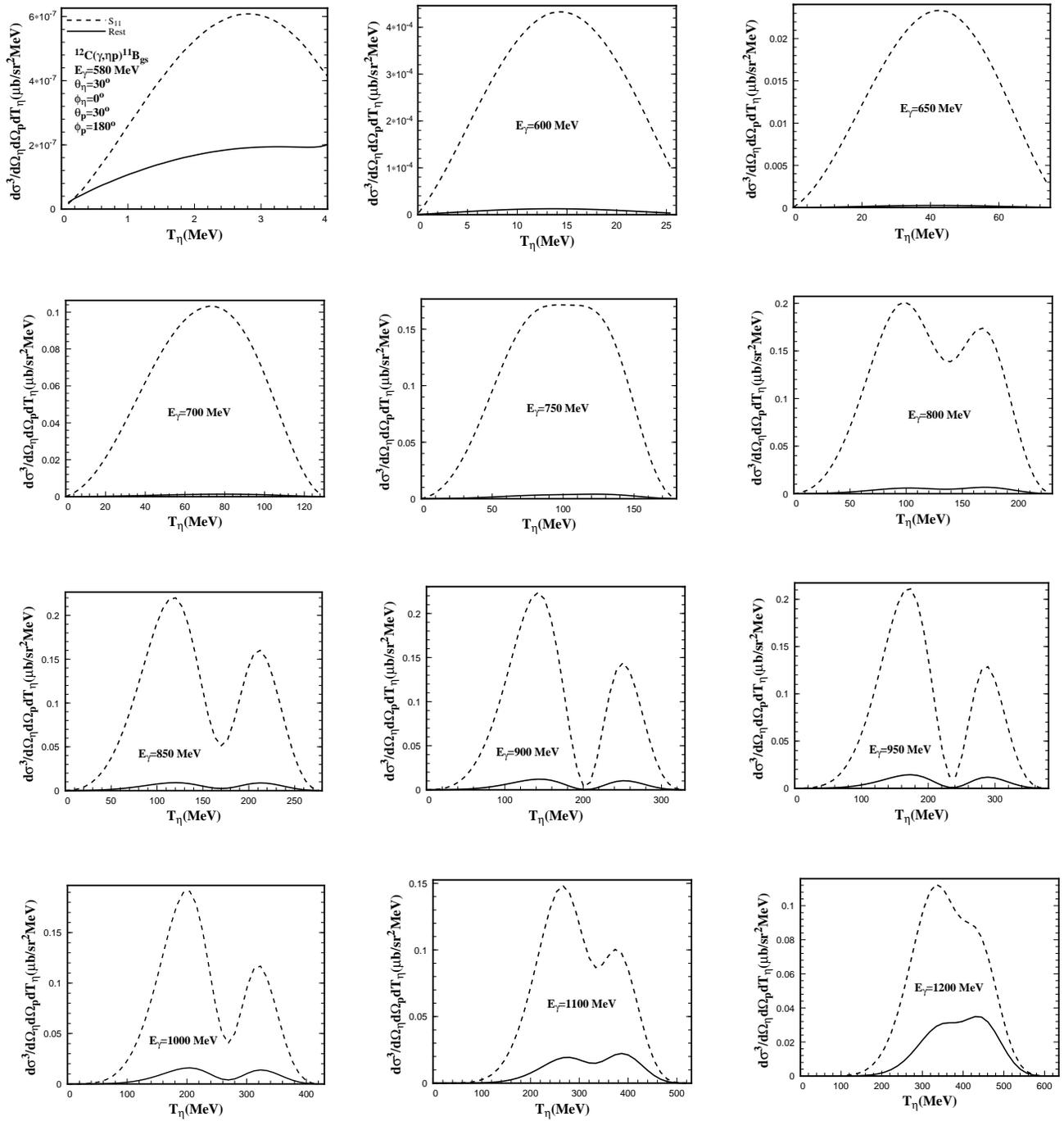
\noindent FIG. 4. The plane wave cross sections for the $^{12}C(~\gamma, \eta p~)^{11}B_{gs}$ 
reaction for photon energies from near threshold up to 1.2 GeV.

\vspace{2.5 mm}
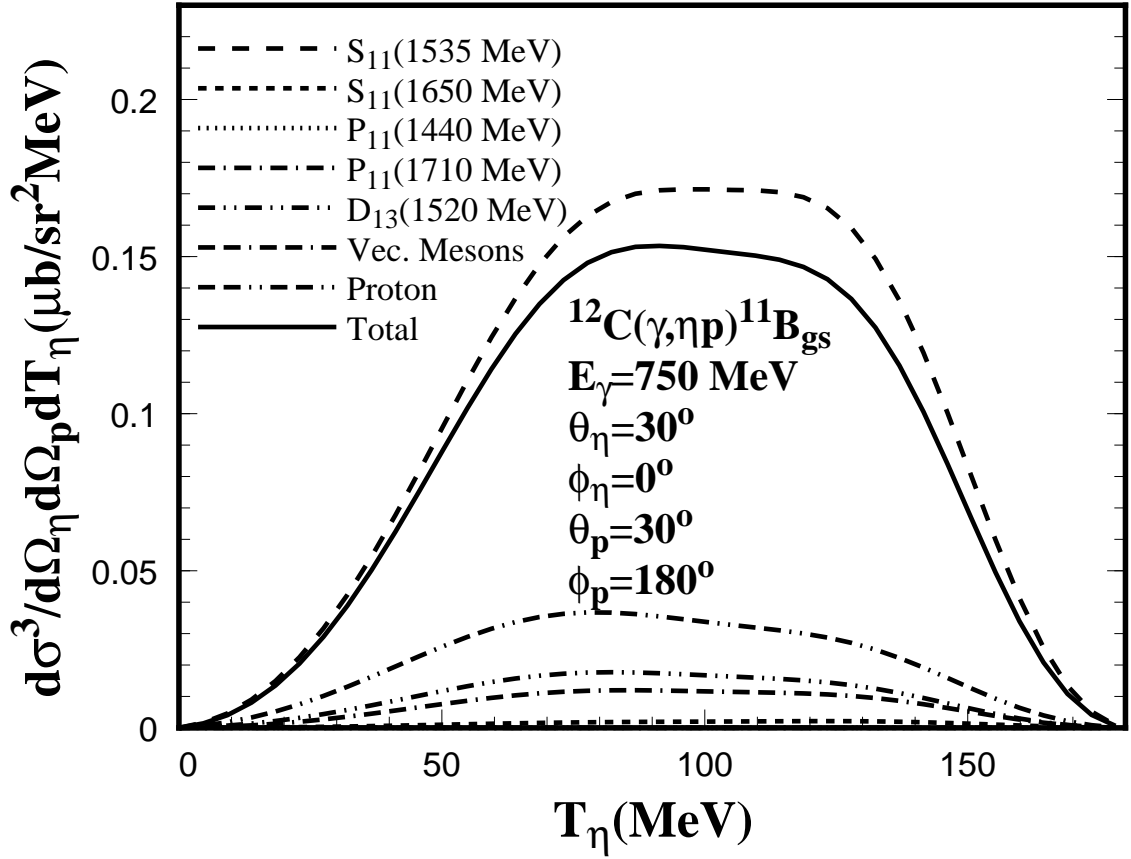
\noindent FIG. 5. Contributions of different diagrams to the cross section of the 
$^{12}C(~\gamma, \eta p~)^{11}B_{gs}$ reaction at photon energies of 750 MeV.

\vspace{2.5 mm}
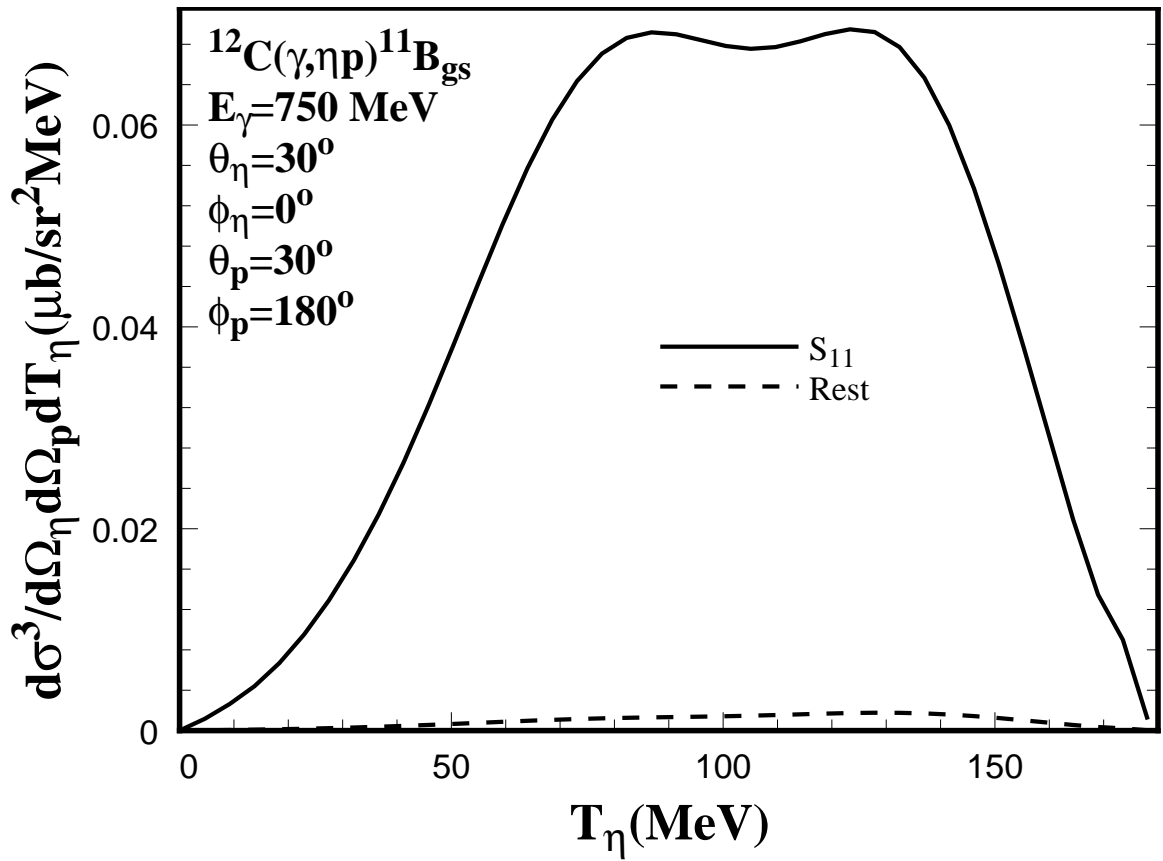
\noindent FIG. 6. Differential cross section for the same reaction as Fig.5, 
calculated in the distorted wave approximation.
Solid curve - DW calculations using only S$_{11}$(1535) diagrams 
( curve labelled S$_{11}$ ).
Long dashed curve - DW calculations using all the diagrams but those of 
S$_{11}$(1535) resonance ( curve labelled Rest ).
 
\vspace{2.5 mm}
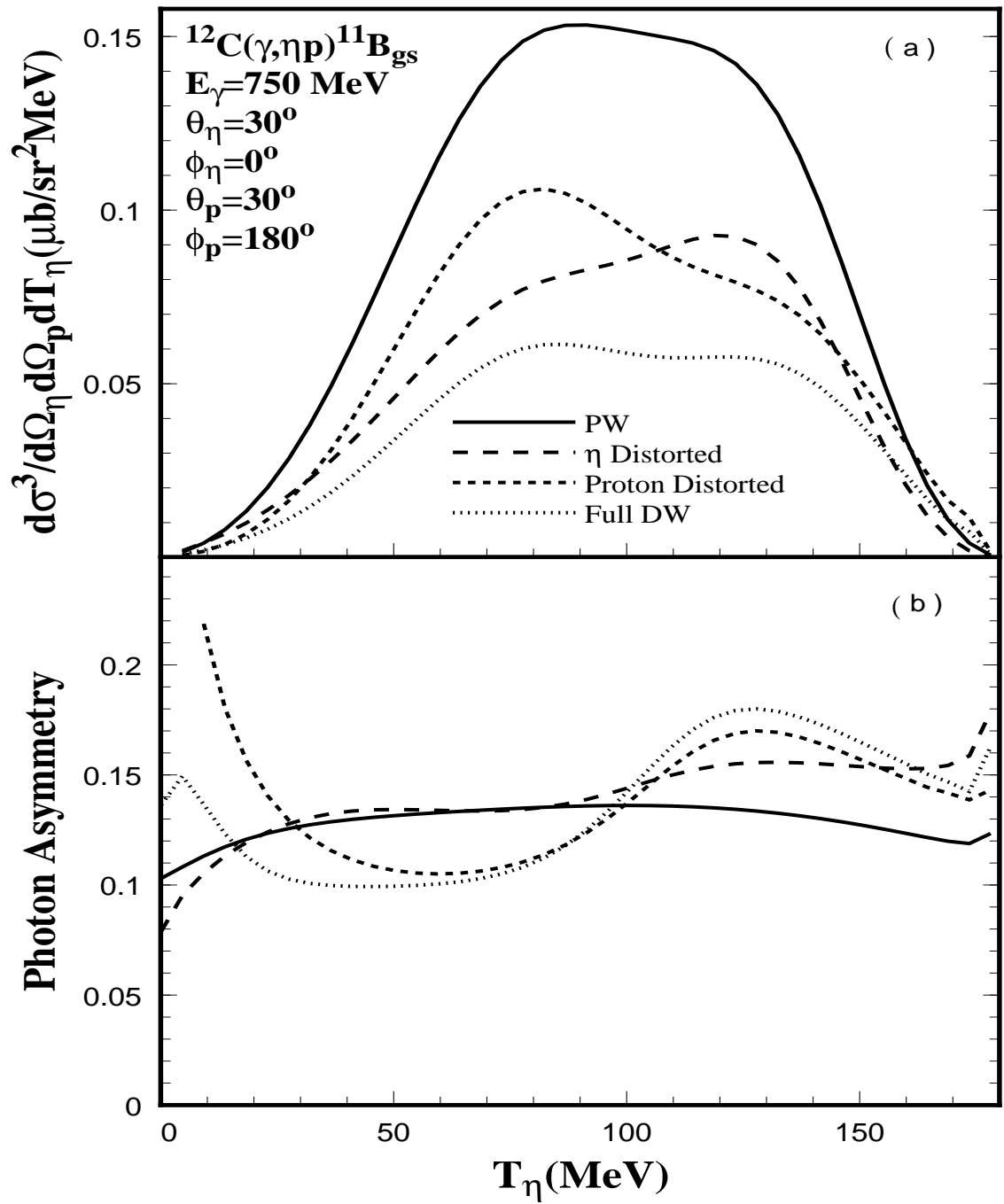
\noindent FIG. 7. Differential cross section (a), and photon 
asymmetry (b) for the reaction  
$^{12}C(\gamma,~\eta p)^{11}B_{g.s.}$ at $E_{\gamma} = 750$ MeV. 
The Hartree potentials of reference \protect\cite{4HOR.NPA} are used in 
calculation of the bound state wavefunction. 
The final state energy dependent global optical potentials are taken from 
reference\protect\cite{4COO.PRC}. 
The $\eta$ optical potential is the DW1 potential of Lee {\it et al.} 
\protect\cite{4LEE.NPA}. 
Solid curve - plane wave calculations. 
Long dashed curve - calculations include only final state 
interactions of the $\eta$ meson (~$\eta$ Distorted~).
Short dashed curve - calculations include only final state 
interactions of outgoing proton (~Proton Distorted~).
Dotted curve - both $\eta$ and proton waves are distorted (~Full DW~).
 
\vspace{2.5 mm}
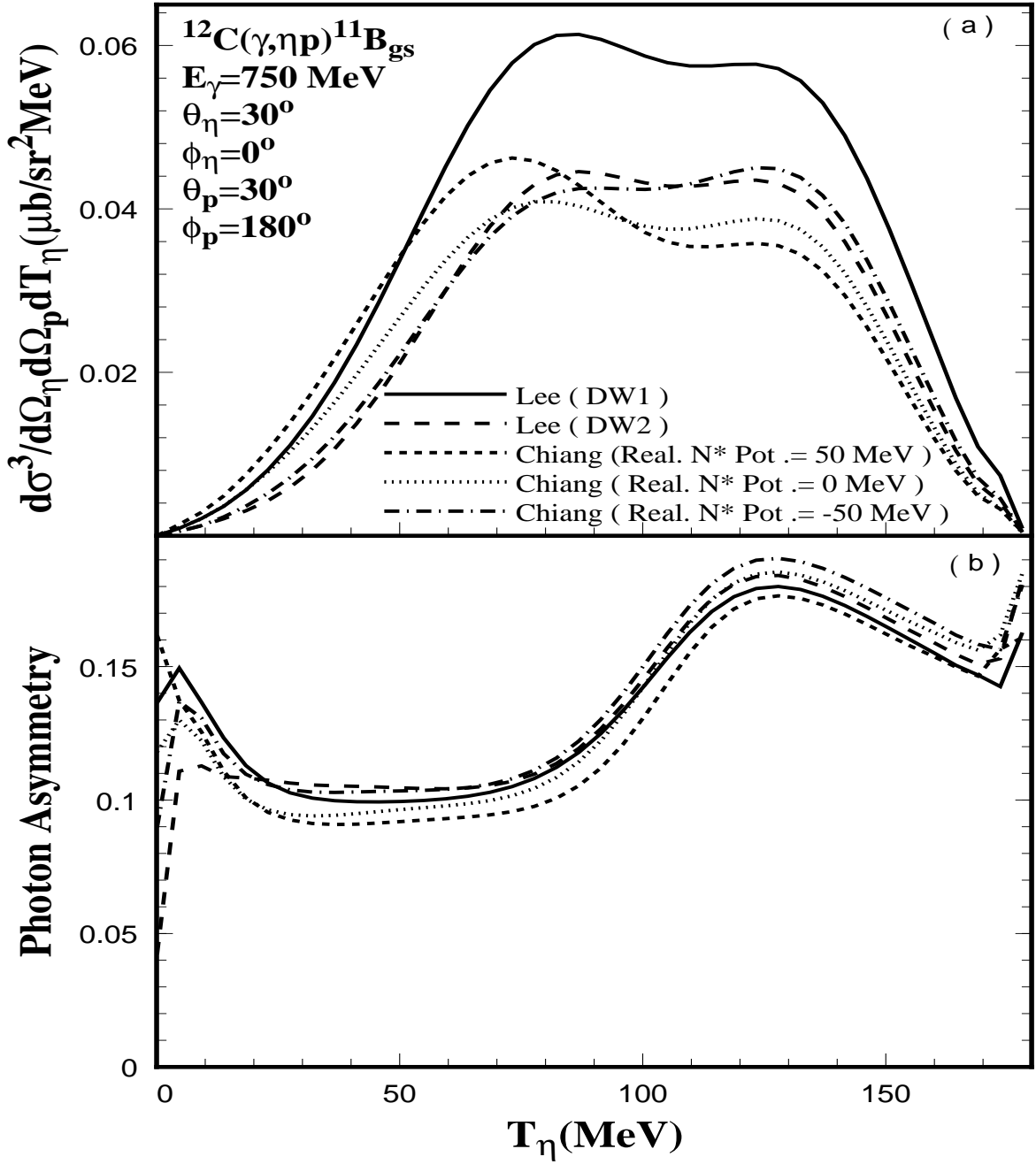
\noindent FIG. 8. Same as Fig.7 but for different choices of $\eta$ 
optical potentials. 
Solid curve - DW calculations using DW1 of reference \protect\cite{4LEE.NPA}. 
Long dashed curve - DW calculations using DW2 of reference 
\protect\cite{4LEE.NPA}. 
DW calculations using the optical potential of reference 
\protect\cite{4OSE.PRC} with the real part of the 
S$_{11}$ self energy 
set to 50 MeV (short dashed curve), 
0 Mev (~dotted curve) and -50 MeV (~dash-dotted curve). 
 
\vspace{2.5 mm}
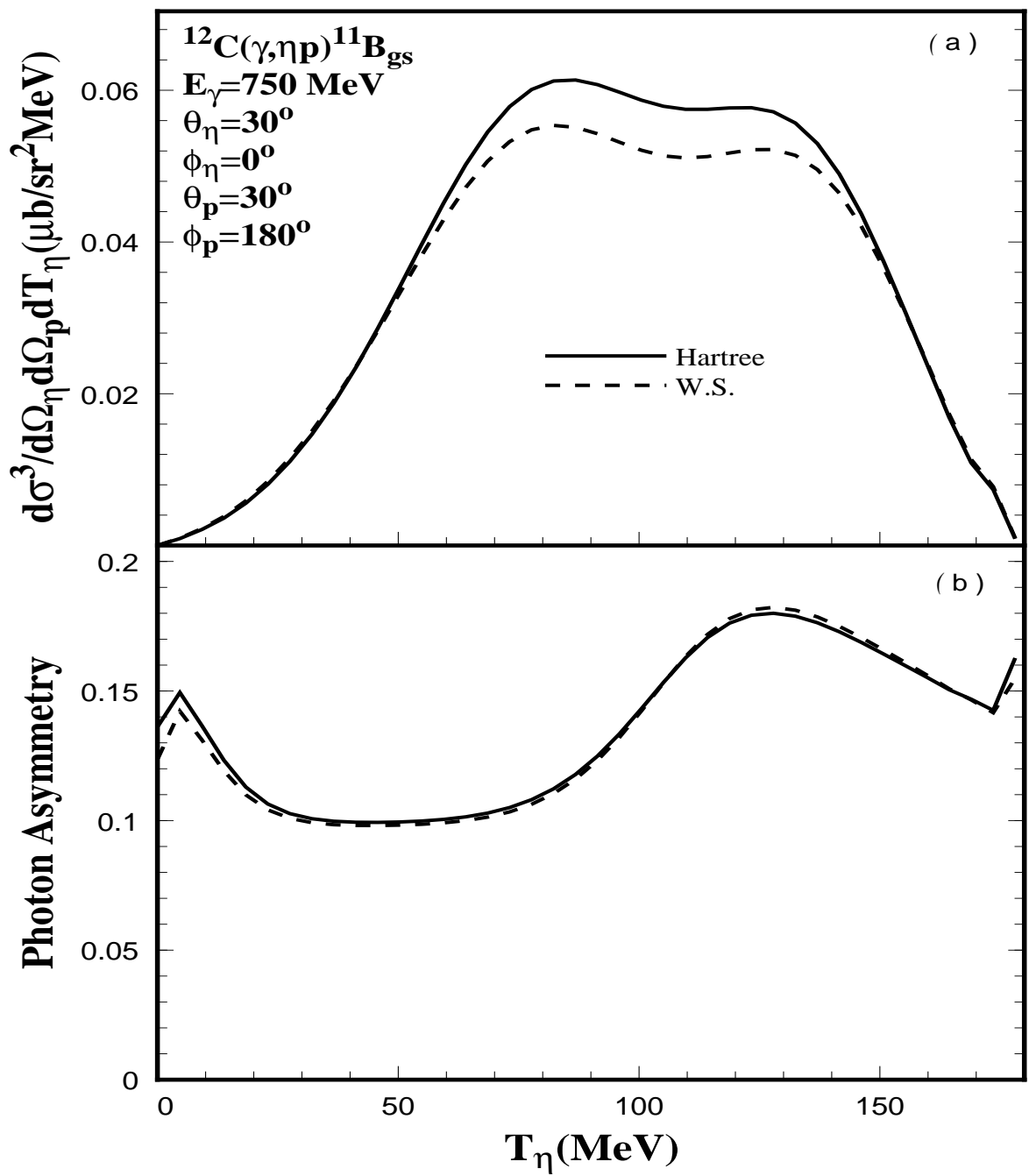
\noindent FIG. 9. Differential cross section (a), and photon 
asymmetry (b) for the same reaction as in Fig.7. 
The final state optical potentials for the $\eta$ meson and outgoing proton 
are the same 
as those of Fig.7. Curves are labelled for different choices 
of proton 
bound potentials: 
Solid curve - DW calculations using Hartree potential of reference 
\protect\cite{4HOR.NPA} and 
dashed curve - DW calculations using Woods-Saxon potential of reference 
\protect\cite{4GER.THS}. 
 
\vspace{2.5 mm}
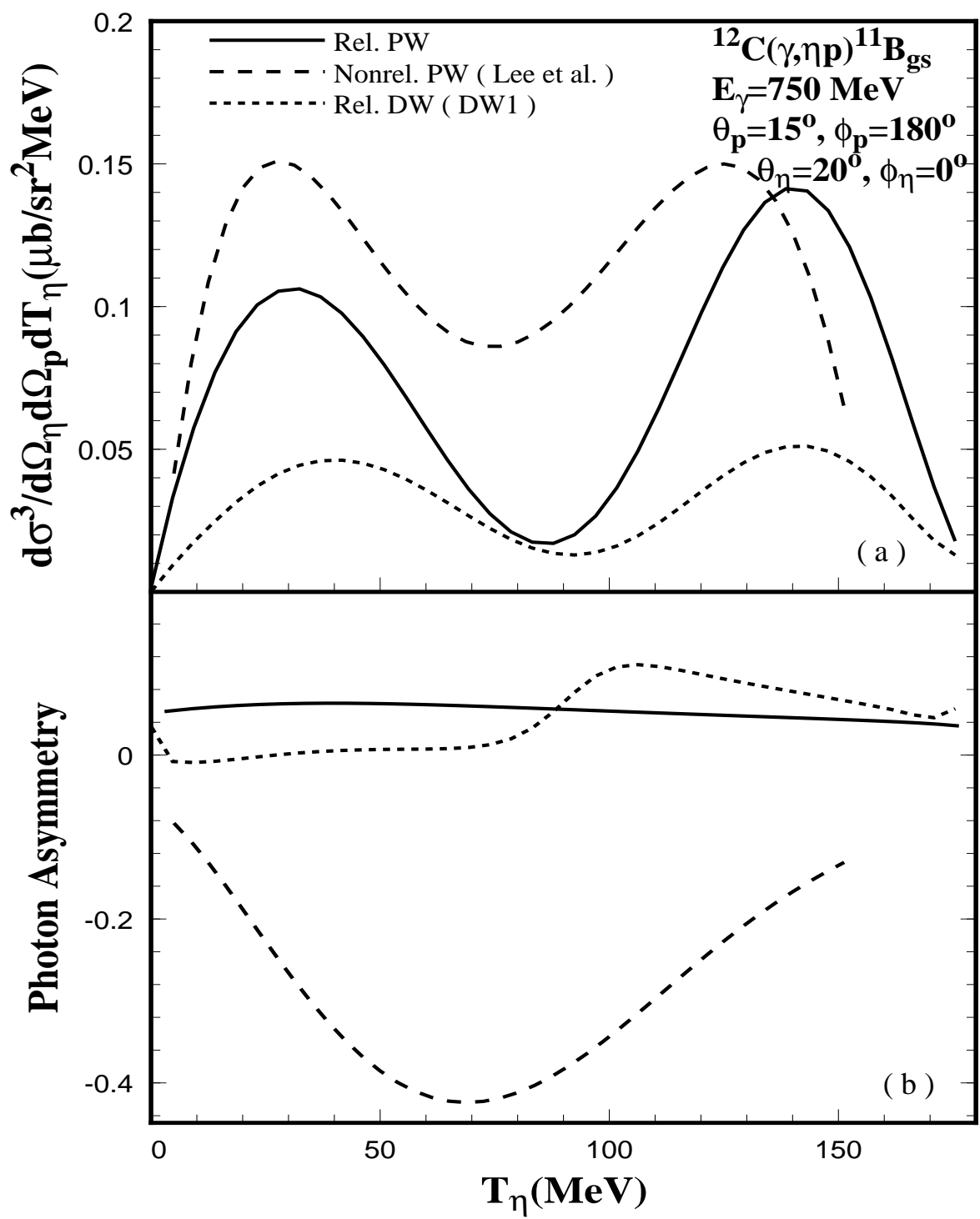
\noindent FIG. 10. Differential cross section (a), and photon 
asymmetry (b) for the same reaction as Fig.7. 
Curves are labelled as: 
Solid curve - relativistic PW calculations.
Long dashed curve - nonrelativistic PW calculations of Lee {\it et al.} \protect\cite{4LEE.NPA}. 
Short dashed curve - relativistic DW calculations.

\vspace{2.5 mm}
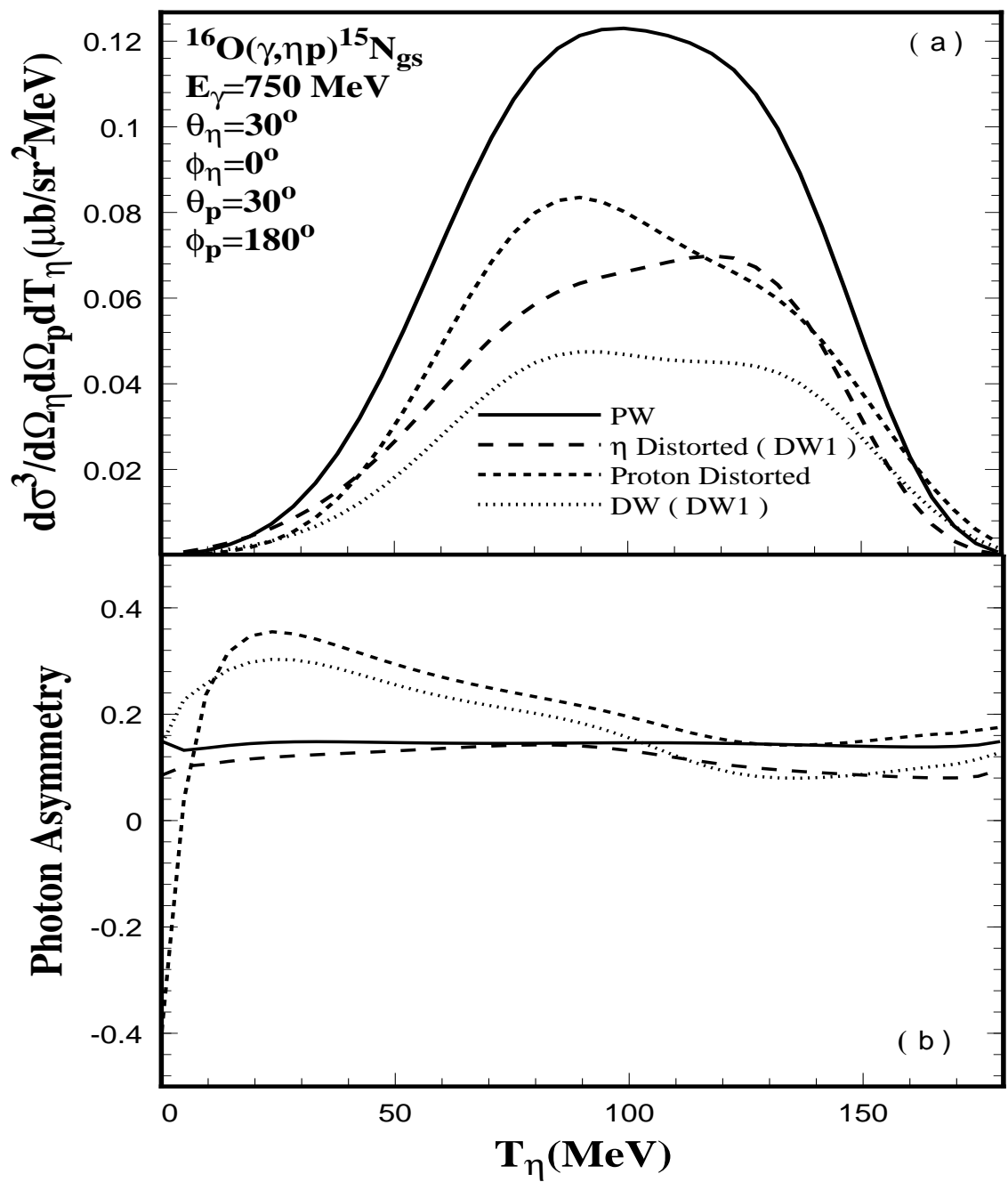
\noindent FIG. 11. Differential cross section (a), and photon 
asymmetry (b) for the reaction  
$^{16}O(\gamma,~\eta p)^{15}N_{g.s.}$ at $E_{\gamma} = 750$ MeV. 
The Hartree potential of reference \protect\cite{4HOR.NPA} is used in 
calculation of 
the bound state wavefunction. 
The final state energy dependent global optical potentials are taken from 
reference\protect\cite{4COO.PRC}. The $\eta$ optical potential is the DW1 
potential 
of Lee {\it et al.} \protect\cite{4LEE.NPA}. 
Solid curve - plane wave calculations. 
Long dashed curve - calculations include only final state 
interactions of the $\eta$ meson with nuclei (~$\eta$ Distorted~).
Short dashed curve - calculations include only final state 
interactions of outgoing proton (~Proton Distorted~).
Dotted curve - both $\eta$ and proton waves are distorted (~DW~). 
  
\vspace{2.5 mm}
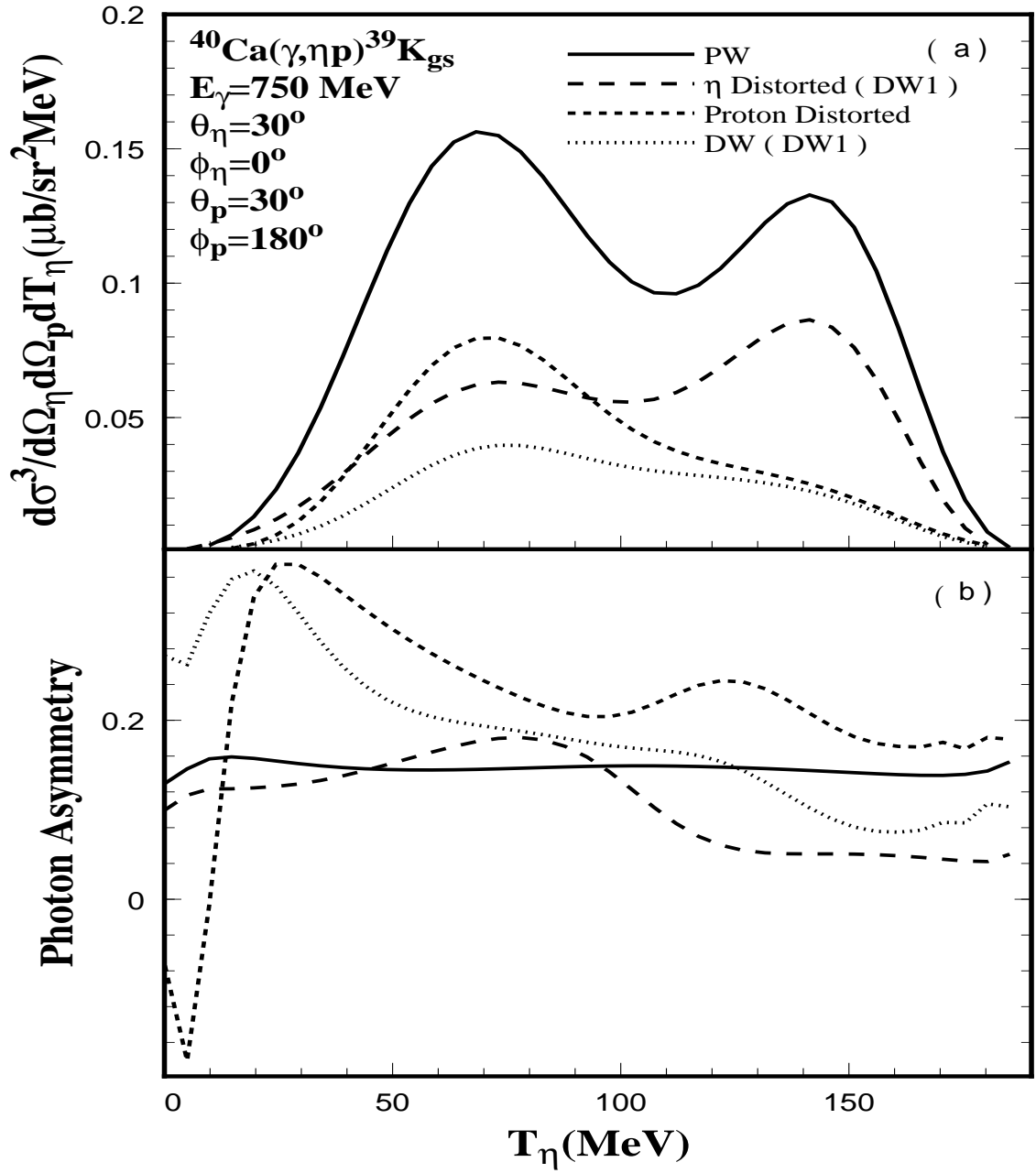
\noindent FIG. 12. Differential cross section (a), and photon 
asymmetry (b) for the reaction  
$^{40}Ca(\gamma,~\eta p)^{39}K_{g.s.}$ at $E_{\gamma} = 750$ MeV. 
The Hartree potential of reference \protect\cite{4HOR.NPA} is used in the 
calculation of 
the bound state wavefunction. 
The final state energy dependent global optical potentials are taken from 
reference\protect\cite{4COO.PRC}. The $\eta$ optical potential is the DW1 
potential 
of Lee {\it et al.} \protect\cite{4LEE.NPA}. 
Solid curve - plane wave calculations. 
Long dashed curve - calculations include only final state 
interactions of the $\eta$ meson with nuclei (~$\eta$ Distorted~).
Short dashed curve - calculations include only final state 
interactions of outgoing proton (~Proton Distorted~).
Dotted curve - both $\eta$ and proton waves are distorted (~DW~). 

\vspace{2.5 mm}
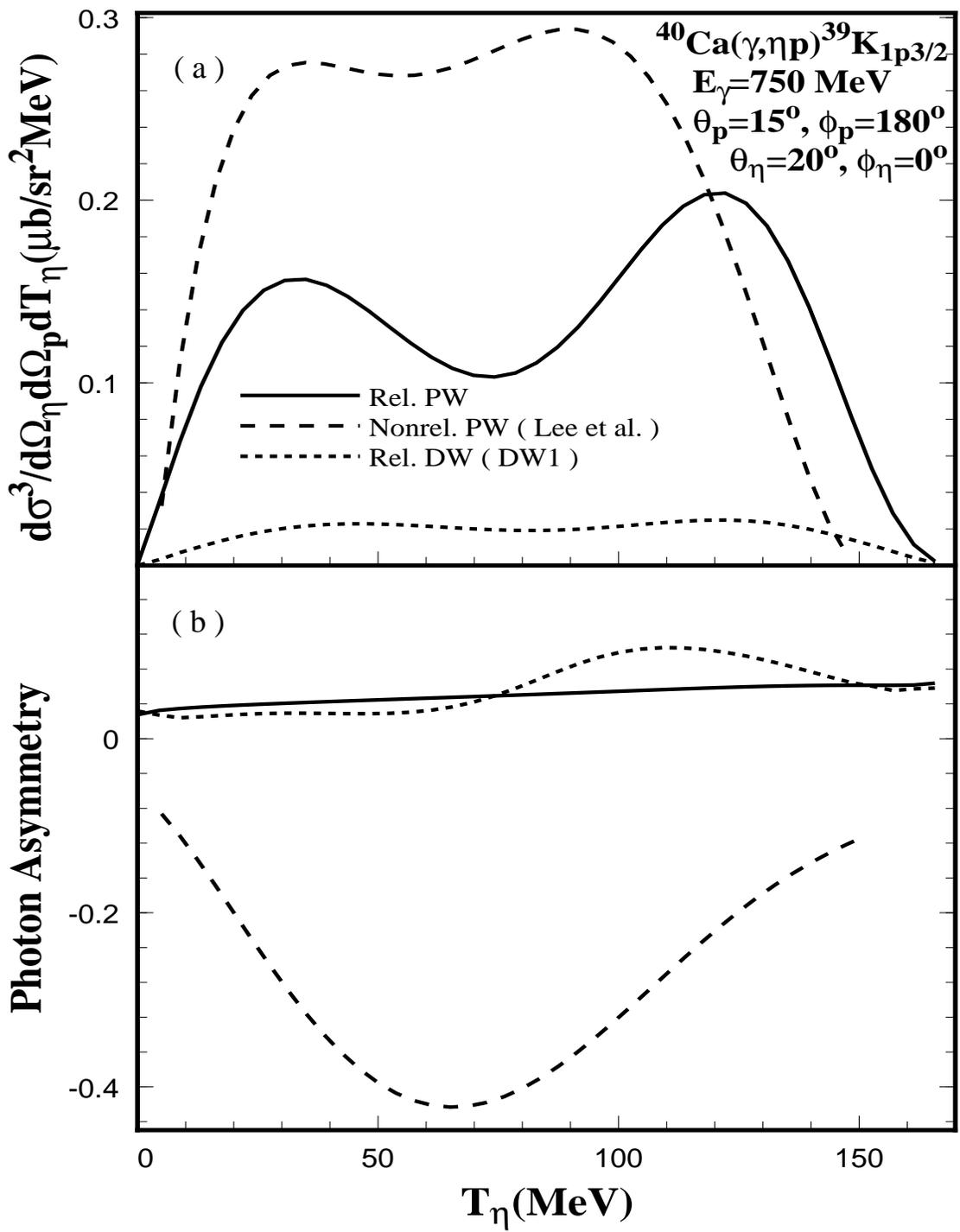
\noindent FIG. 13. Differential cross section (a), and photon 
asymmetry (b) for the same reaction as Fig.12. 
The reaction is assumed to take place in 1p$_{3/2}$ proton. 
The potentials are the same as those of Fig.12.  
Curves are labelled as in Fig.10.
\newpage
\begin{table}
\begin{center}
\begin{tabular}{c c c c c}\hline
Resonance & $J^\pi$ &L$_{\tiny 2I2J}$& Mass(MeV)&$\Gamma$(MeV)  \\ \hline
N*(1440)    & ${1/2}^+$ &P$_{11}$ & 1440&  350           \\ \hline
N*(1520)    & ${3/2}^-$ &D$_{13}$ & 1520&  120           \\ \hline
N*(1535)    & ${1/2}^-$ &S$_{11}$ & 1535&  150           \\ \hline
N*(1650)    & ${1/2}^-$ &S$_{11}$ & 1650&  150           \\ \hline
N*(1710)    & ${1/2}^+$ &P$_{11}$ & 1710&  100           \\ \hline
\end{tabular}
\end{center}
\caption{}
\label{resonances}
\end{table}
\end{document}